\documentclass{aa}
\usepackage{txfonts}
\usepackage{aalongtable}
\usepackage{graphicx}
\usepackage{natbib}
\bibpunct{(}{)}{;}{a}{}{,}

\begin{document}

\title{XMM-{\it Newton} observations of the $\sigma$~Ori cluster. \\
II. Spatial and spectral analysis of the full EPIC field
\thanks{Based on observations obtained with XMM-Newton, an ESA science
mission with instruments and contributions directly funded by ESA Member
States and NASA}}
\author{E. Franciosini\inst{1} \and R. Pallavicini\inst{1} \and 
J. Sanz-Forcada\inst{2}}
\institute{INAF - Osservatorio Astronomico di Palermo,
Piazza del Parlamento 1, I-90134 Palermo, Italy
\and
Astrophysics Division -- Research and Science Support Department
of ESA, ESTEC, Postbus 299, NL-2200 AG Noordwijk, The Netherlands}

\offprints{E. Franciosini, \email{francio@astropa.unipa.it}}

\date{Received 9 June 2005/Accepted 15 September 2005}
\titlerunning{XMM-{\it Newton} observations of the $\sigma$~Ori cluster. II.}

\abstract{
We present the results of an XMM-{\it Newton} observation of the young
($\sim 2-4$ Myr) cluster around the hot star $\sigma$~Orionis. In a previous
paper we presented the analysis of the RGS spectrum of the central hot star;
here we discuss the results of the analysis of the full EPIC field. We have
detected 175 X-ray sources, 88 of which have been identified with cluster
members, including very low-mass stars down to the substellar limit. We
detected eleven new possible candidate members from the 2MASS catalogue. We
find that late-type stars have a median $\log L_\mathrm{X}/L_\mathrm{bol}
\sim -3.3$, i.e. very close to the saturation limit. We detected significant
variability in $\sim 40$\% of late-type members or candidates, including 10
flaring sources; rotational modulation is detected in one K-type star and
possibly in another 3 or 4 stars. Spectral analysis of the brightest sources
shows typical quiescent temperatures in the range $T_1\sim 0.3-0.8$ keV and
$T_2\sim 1-3$ keV, with subsolar abundances $Z\sim 0.1-0.3 \,Z_\odot$,
similar to what is found in other star-forming regions and associations. We
find no significant difference in the spectral properties of classical and
weak-lined T~Tauri stars, although classical T~Tauri stars tend to be less
X-ray luminous than weak-lined T~Tauri stars.

}
\maketitle

\section{Introduction}
\label{intro}

The \object{$\sigma$~Ori cluster} was discovered by {\it ROSAT} \citep{wolk96,
walter97} around the O9.5V binary star \object{$\sigma$~Ori~AB} (which
results from two components separated by 0.2$\arcsec$). It belongs to the
\object{Orion OB1b} association and is located at a distance of
$352_{-85}^{+166}$ pc \citep[from {\it Hipparcos},][]{esa97}. In addition to
several hot stars, it is known to contain more than 100 likely pre-main
sequence (PMS) late-type stars within $30^\prime$ of $\sigma$~Ori
\citep{wolk96,sww04}, as well as several brown dwarfs and planetary-mass
objects \citep{bejar99,bejar01,bejar04,zapat00,caballero04}. The estimated
age of the cluster is $2-4$ Myr \citep{zapat02,oliveira02,sww04}.

We have observed the $\sigma$~Ori cluster using XMM-{\it Newton}. The
observation was centered on the hot star $\sigma$~Ori~AB, in order
to obtain both a high-resolution RGS spectrum of the central source and EPIC
imaging data and low-resolution spectra over the whole field. The analysis
of the RGS and EPIC spectra of the hot star $\sigma$~Ori~AB has
been presented in a previous paper \citep[][hereafter referred to as
Paper~I]{sanz04sig}, together with the analysis of the EPIC spectra of other
three nearby bright sources that could potentially contaminate the RGS
spectrum of $\sigma$~Ori~AB. Of these, only the B2Vp star
\object{$\sigma$~Ori~E} was found to significantly affect the RGS spectrum:
its EPIC spectrum allowed us to correct for its contribution, and therefore
to derive the emission measure distribution and elemental abundances of
$\sigma$~Ori~AB. We found that $\sigma$~Ori~AB has a much softer spectrum
than the other sources, consistently with a wind origin, however the RGS
spectrum shows no evidence for line broadenings and shifts with velocities
$\ga 800$~km~s$^{-1}$, as could be produced by strong winds; the low $f/i$
line ratio in the Ne and O He-like triplets indicates either high density
(i.e. magnetic confinement either close to or far from the star) or, more
likely, a strong UV radiation field (i.e. emission close to the star where
the wind is however too weak to produce shocks). 

In Paper~I we also reported the detection of a strong flare from the
magnetic hot star $\sigma$~Ori~E, which is not expected from models of X-ray
emission from winds. Based on the characteristics of the flare and of the
quiescent emission, whose spectrum is harder than that of $\sigma$~Ori~AB
and consistent with those of late-type stars, we argued that the flare and
most of the quiescent emission is most likely due to an unseen late-type
companion, although emission from the magnetic hot star itself cannot be
excluded (see discussion in Paper~I).

In this paper we present the analysis of the full EPIC field, in order to
derive the X-ray properties of the other members or candidates of the
$\sigma$~Ori cluster. The paper is organized as follows. X-ray observations
and data analysis are described in Sect.~\ref{observ}. In
Sect.~\ref{results} we discuss the X-ray properties of cluster members and
compare our results with those of other star forming regions and young open
clusters. Spectral analysis of the brightest sources is presented in
Sect.~\ref{spectra}. Discussion and conclusions are given in
Sect.~\ref{concl}.

\begin{figure}
\resizebox{\hsize}{!}{\includegraphics[clip]{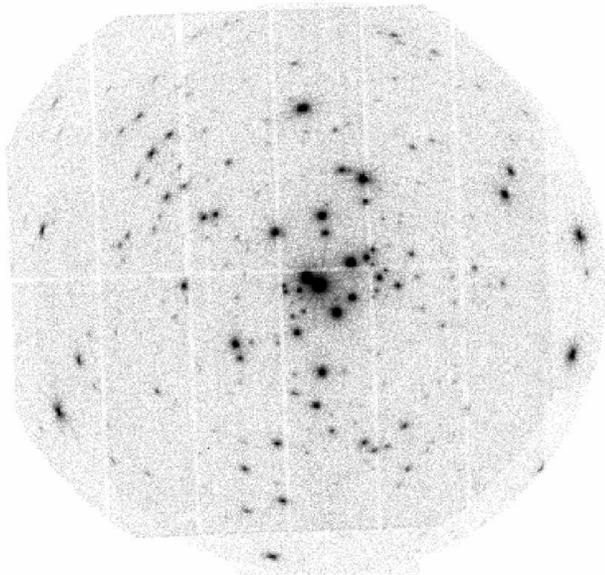}}
\caption{Composite EPIC (MOS1+MOS2+PN) image of the $\sigma$~Ori cluster}
\label{sori_ima}
\end{figure}

\section{Observations and data analysis}
\label{observ}

XMM-{\it Newton} observations of the $\sigma$~Ori cluster, centered on the
hot star $\sigma$~Ori~AB, were carried out as part of the Guaranteed Time of
one of us (R.P.) using both the EPIC MOS and PN cameras and the RGS
instrument. The observation (ID 0101440301) started at 21:47 UT on March 23,
2002 and ended at 9:58 UT on March 24, 2002, for a total duration of 43 ks.
The EPIC cameras were operated in Full Frame mode using the thick filter.

Data analysis was carried out using the standard tasks in SAS v.5.4.1. The
analysis of the RGS data has been discussed in Paper~I and will not be
repeated here. EPIC calibrated and cleaned event files were derived from the
raw data using the standard pipeline tasks {\sc emchain} and {\sc epchain}
and then applying the appropriate filters to eliminate noise and bad events.
The event files have also been time filtered in order to exclude a few short
periods of high background due to proton flares; the final effective
exposure time is 41 ks for each MOS and 36 ks for the PN. We limited our
analysis to the $0.3-7.8$ keV energy band, since events below 0.3 keV are
mostly unrelated to bona-fide X-rays, while above 7.8 keV only background is
present. The combined EPIC (MOS1+MOS2+PN) image in the $0.3-7.8$ keV energy
band is shown in Fig.~\ref{sori_ima}. Exposure maps for each instrument in
the same energy band were created using the task {\sc eexpmap}.

Source detection was performed both on the individual datasets and on the
merged MOS1+MOS2+PN dataset using the Wavelet Detection algorithm developed
at INAF-Osservatorio Astronomico di Palermo \citep[Damiani et al., in
preparation]{damiani97}, adapted to the EPIC case. From the comparison of
the count rates of common sources detected on the individual datasets, we
derived a median ratio of PN to MOS count rates of $\sim 3.2$; this value
was then used as a scaling factor for the PN exposure map in the detection
on the summed dataset, in order to take into account the different
sensitivities of the PN and MOS cameras. This implies a resulting MOS
equivalent exposure time of $\sim 200$ ks for the merged dataset. Count
rates derived from the detection on the summed dataset are MOS equivalent
count rates.

We used a significance detection threshold of $5\sigma$, chosen in order to
have at most one spurious detection, and determined from a set of 100
simulations of pure background datasets with the same number of counts as
the observation. After removing a few obviously spurious detections (due to
hot pixels, to the point spread function structure of the central bright
source, and to out-of-time events) we obtained a total of 175 sources, three
of which were detected only on a single instrument dataset (two on MOS and
one on PN). 

\subsection{The optical catalogue}
\label{catalog}

The $\sigma$~Ori cluster has been the subject of several studies after its
discovery by {\it ROSAT} \citep{walter97}. \citet{wolk96} performed
follow-up UBVRI photometry down to $V\sim 18$ within 30$\arcmin$ from
$\sigma$~Ori, finding $\sim 130$ candidate PMS stars, and obtained
medium-resolution spectroscopy for a few of them. Recently, \citet{sww04}
extended the survey to an area of $\sim 0.89$~deg$^2$ around the hot star
down to $V\sim 20$. Several deep optical and infrared surveys of the
cluster, with limiting magnitude $I\sim 22-24$, have led to the discovery of
several very-low mass stars and brown dwarfs as well as of a number of
planetary-mass objects \citep{bejar99,bejar01,bejar04,zapat00,caballero04}.
Additional photometric RI surveys down to $I\sim 18-19$ have been performed
by \citet{se04} and \citet{kenyon05}. Low- and medium-resolution
spectroscopic observations are available for a subsample of low-mass stars
and brown dwarfs
\citep{bejar99,barrado01,barrado03,martin01,zapat02,kenyon05,burn05}.

\begin{figure}
\resizebox{\hsize}{!}{\includegraphics[clip]{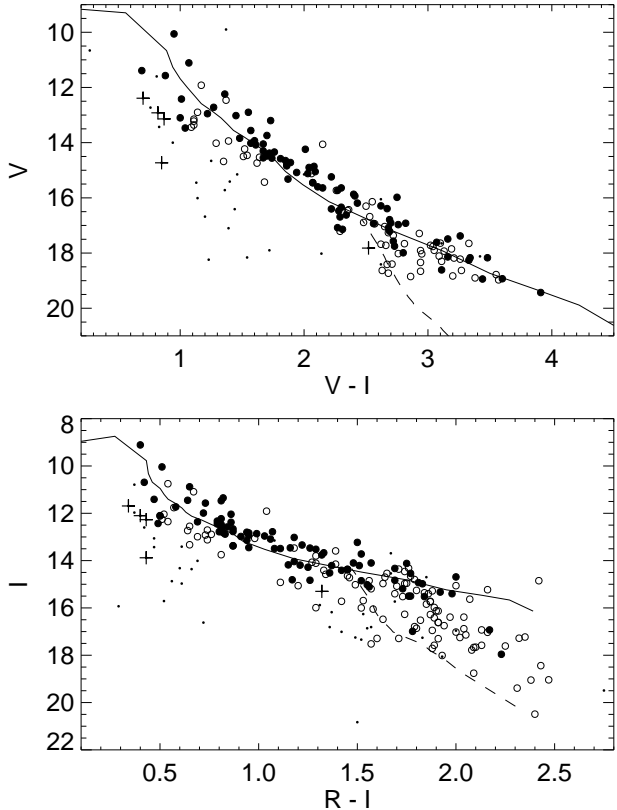}}
\caption{$V$ vs. $(V-I)$ and $I$ vs. $(R-I)$ colour-magnitude diagrams for
stars in the XMM-{\it Newton} field of view. Detected cluster members and
candidates are marked with filled circles, while open circles indicate non
detections. Crosses and dots mark the position of detected and undetected
non-members, respectively. The solid and dashed lines are the 5 Myr
isochrones from \citet{siess00} and \citet{baraffe98}, respectively, shifted
for the distance and reddening of the $\sigma$~Ori cluster}
\label{cmd}
\end{figure}

We constructed our optical catalogue by including all stars from the
above-mentioned studies, plus a few additional H$\alpha$ emission line stars
from \citet{weaver04} and the hot stars from the studies of the Orion OB
association by \cite{wh77} and \citet{brown94}. After cross-correlating the
data from the available studies, we obtain a total of 266 stars falling in
the XMM field of view. About 40\% of the late-type stars in the catalogue
have spectroscopic membership information based on radial velocity, lithium
or other youthness indicators. For the remaining stars only photometry or at
most H$\alpha$ measurements are avalaible: in these cases, we have assigned
a photometric membership based on their position in different
colour-magnitude diagrams, rejecting as members those stars falling below the
10 Myr isochrone in at least two diagrams. For stars without a
spectroscopically-determined spectral type, we derived spectral types from
their $R-I$, $V-I$ or $I-J$ colours using the transformations by \citet{kh95}
and \citet{leggett01}. In total, there are 218 probable or candidate members
in our catalogue, including 8 early-type (O-B-A) stars, two late F-G type
stars, 143 K- and early M-type stars ($\la\,$M4), and 65 very-low mass stars
and brown dwarfs with spectral type later than $\sim\,$M5\footnote{Note
that, at the age of $\sigma$~Ori, the substellar limit lies around spectral
type $\sim\,$M5--M6, depending on the adopted model isochrones \citep[see
e.g.][]{zapat02}.}

Most of the stars in our optical catalogue have counterparts in the 2MASS
All-Sky Point Source Catalogue \citep{2mass}: for these stars we used their
2MASS coordinates in order to have accurate positions.

\begin{figure}
\resizebox{\hsize}{!}{\includegraphics[clip]{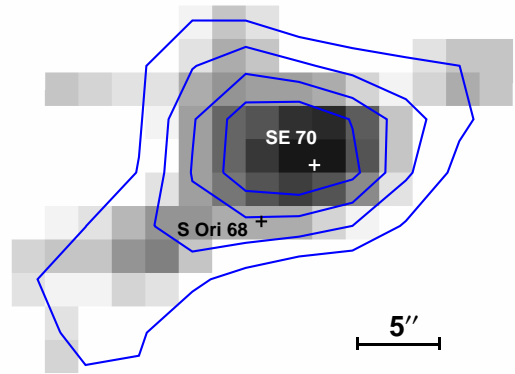}}
\caption{Close-up of the EPIC summed image at the position of the X-ray
source identified with S\,Ori~68 and SE\,70. The positions of the two stars
are marked with crosses. Contour levels are at 0.2, 0.3, 0.4, 0.5
cts~arcsec$^{-2}$}
\label{sori68}
\end{figure}

\begin{table*}
\caption{X-ray sources identified with possible cluster candidates from
2MASS. NX is a running identification number for the X-ray sources. The
column labeled ``Sign.'' indicates the significance of detection}
\begin{tabular}{rccrrcrcc}
\hline\hline
\noalign{\smallskip}
NX& RA$_\mathrm{X}$& DEC$_\mathrm{X}$& Sign.& Count rate& 2MASS& $\Delta r$&
$J$& $J-K$ \\
&\multicolumn{2}{c}{(J2000)}& & (cts/ks)& & ($\,\arcsec\,$)& & \\
\noalign{\smallskip} \hline \noalign{\smallskip}
 14& 5:38:15.63& $-$2:42:05.5& 12.9& $2.67 \pm 0.39$& J05381552-0242051& 0.36& 16.33&     \\
 42& 5:38:31.07& $-$2:34:02.8&  7.0& $0.64 \pm 0.13$& J05383098-0234038& 1.00& 14.92& 0.91\\
 50& 5:38:33.05& $-$2:39:27.2&  8.2& $0.53 \pm 0.11$& J05383302-0239279& 1.30& 14.59& 0.89\\
 91& 5:38:48.47& $-$2:36:42.3&  6.6& $0.74 \pm 0.14$& J05384828-0236409& 1.86& 12.04& 0.90\\
101& 5:38:50.97& $-$2:27:44.4&  6.4& $0.70 \pm 0.17$& J05385101-0227456& 2.35& 14.28& 0.95\\
103& 5:38:51.80& $-$2:36:02.6& 28.2& $3.50 \pm 0.25$& J05385173-0236033& 0.84& 12.91& 0.88\\
116& 5:38:59.13& $-$2:34:15.5&  7.7& $0.63 \pm 0.13$& J05385884-0234131& 3.71& 15.86& 0.97\\
148& 5:39:13.58& $-$2:37:37.8& 10.4& $1.18 \pm 0.18$& J05391346-0237391& 1.31& 13.41& 0.91\\
151& 5:39:15.89& $-$2:36:49.7&  7.4& $0.74 \pm 0.15$& J05391582-0236507& 1.15& 13.25& 1.03\\
153& 5:39:17.12& $-$2:41:17.6& 22.7& $4.20 \pm 0.37$& J05391699-0241171& 0.59& 14.29& 0.92\\
175& 5:39:40.24& $-$2:43:07.9&  7.5& $1.94 \pm 0.46$& J05393998-0243097& 3.04& 10.65& 1.12\\
\noalign{\smallskip}\hline
\end{tabular}
\label{detcand}
\end{table*}

\begin{table*}
\caption{X-ray sources identified with probable cluster non-members or with
stars without membership information. NX is a running identification number
for the X-ray sources. The column labeled ``Sign.'' indicates the
significance of detection. Optical identifications labeled 4771-... and
r05... are from \citet{wolk96}; SWW\,222 is from \citet{sww04}}
\begin{tabular}{rccrrlcccc}
\hline\hline
\noalign{\smallskip}
NX& RA$_\mathrm{X}$& DEC$_\mathrm{X}$& Sign.& Count rate& Optical ID&
$\Delta r$& $I$& $R-I$& Notes \\
&\multicolumn{2}{c}{(J2000)}& & (cts/ks)& & ($\,\arcsec\,$)& & & \\
\noalign{\smallskip} \hline \noalign{\smallskip}
 41& 5:38:30.06& $-$2:23:36.6&  26.1& $ 7.50 \pm 0.59$& r053829-0223           & 1.66& 12.27& 0.43& $a$  \\
 46& 5:38:32.21& $-$2:32:43.6&  13.7& $ 1.14 \pm 0.15$& GSC2 S02003215575      & 0.62&      &     &      \\
 47& 5:38:32.70& $-$2:31:16.1&   6.4& $ 0.57 \pm 0.13$& 2MASS J05383268-0231156& 1.29&      &     &      \\
 56& 5:38:34.30& $-$2:34:59.9&  21.9& $ 3.52 \pm 0.35$& r053834-0234           & 1.63& 13.88& 0.43& $a$  \\
114& 5:38:57.20& $-$2:31:25.5&   7.2& $ 0.65 \pm 0.13$& 4771-1071              & 1.74& 12.10& 0.40& $a$  \\
119& 5:38:59.61& $-$2:45:08.6&  38.3& $ 9.36 \pm 0.53$& 4771-0026              & 0.68& 11.69& 0.34& $a$  \\
120& 5:38:59.65& $-$2:35:27.9&   9.1& $ 0.76 \pm 0.16$& 2MASS J05385930-0235282& 3.83&      &     &      \\
123& 5:39:01.19& $-$2:33:37.8&  13.8& $ 1.42 \pm 0.18$& GSC2 S02003215312      & 4.70&      &     &      \\
125& 5:39:01.58& $-$2:38:57.0& 132.8& $45.72 \pm 0.91$& HD 37525               & 0.67&  8.07&     & $b$  \\
169& 5:39:30.72& $-$2:38:28.7&  10.1& $ 1.77 \pm 0.29$& SWW 222                & 1.93& 15.30& 1.32& $c$  \\
\noalign{\smallskip} \hline \noalign{\smallskip}
\multicolumn{10}{l}{$a$: probable non member on the basis of photometry}\\
\multicolumn{10}{l}{$b$: non member from proper motion \citep{wh77}}\\
\multicolumn{10}{l}{$c$: non member from radial velocity \citep{burn05}}\\
\end{tabular}
\label{detnm} 
\end{table*}

\subsection{Source identification}
\label{srcid}

In order to find optical counterparts to our X-ray source list, we have
determined the optimal search radius by constructing the cumulative
distribution of the offsets between X-ray and optical position, following
\citet{randsch95}. We chose a search radius of $5\arcsec$, for which less
than two spurious identifications are expected. After correlating the X-ray
source list with the optical catalogue, we found a median offset of $\sim
1.4\arcsec$ in right ascension between the X-ray and optical positions; we
therefore corrected the X-ray positions, and repeated the identification
process. We found 88 sources with at least one cluster member or candidate
within $5\arcsec$ (including two double identifications); these sources are
listed in Table~\ref{detmem} together with their optical properties. The
position of detected members in the $V-(V-I)$ and $I-(R-I)$ colour-magnitude
diagrams are shown in Fig.~\ref{cmd}. X-ray detections delineate very clearly
the cluster sequence, confirming the importance of X-ray emission as a
membership indicator for young stars.

Of the detected members, three are OB stars, i.e. $\sigma$~Ori~AB (O9.5V),
$\sigma$~Ori~E (B2Vp) and \object{HD~294272} (B9.5III), and one is an A8V
star (\object{HD~37564}). We have detected the two FG-type stars, $\sim
64$\% of K-type stars and $\sim 50$\% of early-M stars ($<$\,M5). Seven
sources ($\sim 11$\%) have been identified with very low-mass stars and
brown dwarfs of spectral type later than $\sim\,$M5. However, in two cases,
the identification is ambiguous, since there is another early-M star falling
inside the identification radius. One of these cases is source NX~67, which
is identified with \object{SE\,70} ($\sim$\,M4) at $2.6\arcsec$, and
\object{S\,Ori~68}, a planetary-mass object of spectral type L5.0
\citep{bejar01}, at a distance of $4.7\arcsec$. This source underwent a
flare during the observation, with an increase in the count rate by a factor
of $\sim 4$ and a total duration of $\sim 10$ ks (see Sect.~\ref{var},
Fig.~\ref{flares}). Inspection of the X-ray image (Fig.~\ref{sori68}) shows
that the bulk of the X-ray emission, and therefore the flare, is associated
with SE\,70, although we cannot exclude the presence of a very weak
contribution from S\,Ori~68 itself. Unfortunately the statistics outside of
the flare is too low to allow us to derive any information on the relative
contribution of the two objects to the quiescent X-ray source. In the
analysis of Sect.~\ref{Lx} we have therefore assigned all the X-ray flux to
SE\,70, while taking for S\,Ori~68 the same value as upper limit to its
emission.

The other source with two possible counterparts is NX~167, which has been
identified with \object{S\,Ori~J053926.8-022614}, a $\sim$\,M6 star, at
3.9$\arcsec$, and \object{SE\,94} ($\sim$\,M2) at 3.2$\arcsec$. In this
case, it is not possible to identify which star is the most probable X-ray
emitting one; we have therefore equally divided the X-ray flux between them.

The faintest star with a certain X-ray detection is the candidate brown
dwarf \object{S\,Ori~25}, which has a spectral type M7.5 \citep{barrado03}
and an estimated mass of $\sim 0.04\, M_\odot$ \citep{muzerolle03}. Its
X-ray luminosity in the 0.3--8 keV band is $L_\mathrm{X} \sim 3\times
10^{28}$ erg~s$^{-1}$ (see Sect.~\ref{Lx}), corresponding to $\log
L_\mathrm{X}/L_\mathrm{bol} \sim -3.3$, i.e. close to the saturation limit.

For the remaining 124 late-type (F to M) cluster members and candidates with
no associated X-ray source we determined 3$\sigma$ upper limits at the
optical positions using the Wavelet algorithm. Their X-ray and optical
properties are given in Table~\ref{upplim}. 

Six additional X-ray sources have been identified with probable cluster non
members from our optical catalogue, including an early-type star
(\object{HD\,37525}, B5V) which was rejected as member of the OB association
by \citet{wh77} on the basis of proper motion. Interestingly, this source
underwent a flare at the end of the observation, after $\sim 10$ hrs of
quiescent emission; an analysis of its quiescent PN spectrum shows high
temperatures (0.6 and 1.3 keV), similar to those found for the B2Vp star
$\sigma$~Ori\,E (Paper~I) and for late-type stars (see Sect.~\ref{spectra}),
suggesting that the emission might originate from an unseen late-type
companion. Of the other non members, one (\object{SWW\,222}) has a radial
velocity inconsistent with cluster membership \citep{burn05}, while the
others have been rejected on the basis of photometry only. These X-ray
sources are listed in Table~\ref{detnm}.

In order to increase the number of identifications, we have also
cross-correlated the X-ray source list with the 2MASS and the GSC2.2
catalogues, finding 13 counterparts from 2MASS and 2 from GSC2. Eleven of
the 2MASS counterparts have $JHK$ photometry consistent with cluster
membership, and are therefore considered as possible candidate members of
the cluster; we list them in Table~\ref{detcand}. The other identified X-ray
sources are given in Table~\ref{detnm}. 

For the remaining 66 sources (listed in Table~\ref{unid}) we did not find
any known counterpart in any astronomical catalogue. This number is
consistent with the expected number of extragalactic X-ray sources ($\sim
60-80$) in the direction of $\sigma$~Ori, derived using the $\log N-\log S$
relations by \citet{hasinger01} and \citet{tozzi01}. 

\begin{figure}
\resizebox{\hsize}{!}{\includegraphics[clip]{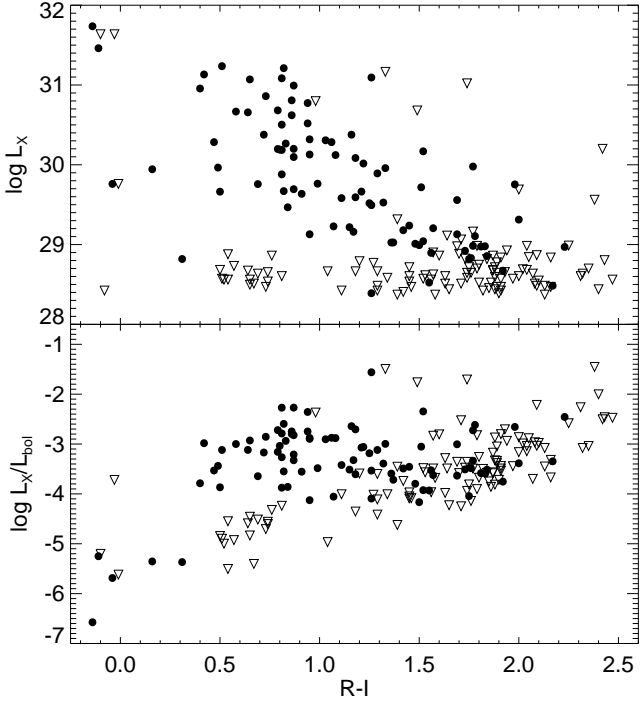}}
\caption{$L_\mathrm{X}$ ({\em top}) and $L_\mathrm{X}/L_\mathrm{bol}$ ({\em
bottom}) as a function of the $(R-I)$ colour for detected (dots) and
undetected (open triangles) cluster members and candidates}
\label{lx-col}
\end{figure}

\begin{figure}
\resizebox{\hsize}{!}{\includegraphics[clip]{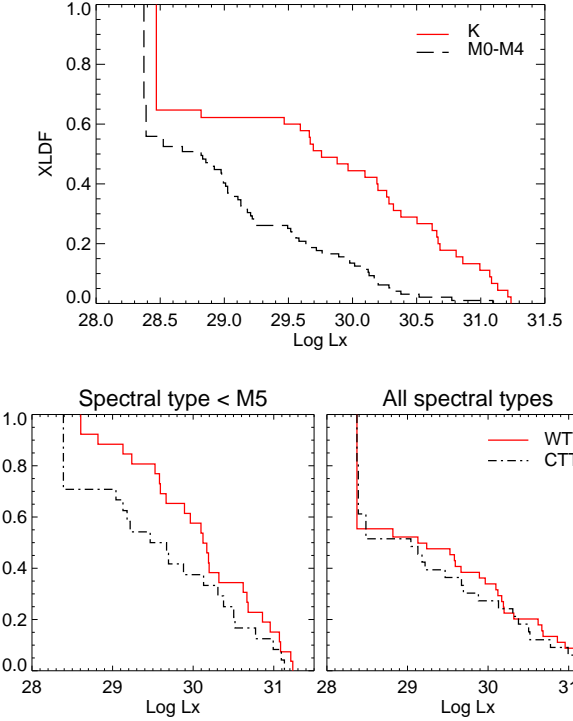}}
\caption{{\em Upper panel}: comparison of the XLDF for K-type (red solid
line) and early M-type (black dashed line) stars in $\sigma$~Ori. 
{\em Lower panels}: comparison of the XLDFs for WTTS (red solid line) and
CTTS (black dot-dashed line), considering only stars earlier than $\sim$\,M5
({\em left}) and all stars and candidate brown dwarfs ({\em right})}
\label{xldf}
\end{figure}

\section{Results}
\label{results}

\subsection{X-ray luminosities}
\label{Lx}

In order to derive X-ray luminosities, we have computed a conversion factor
in the 0.3--8~keV band using the results of the PN and MOS1 spectral fits of
late-type stars (see Sect.~\ref{spectra}). The conversion factor has been
determined by comparing the count rates obtained from the Wavelet algorithm
on the summed dataset with the unabsorbed X-ray flux derived from the
best-fit models, excluding the few sources with absorption higher than the
mean value $N_H = 2.7\times 10^{20}$ cm$^{-2}$, derived from the mean
reddening $E(B-V)=0.05$ \citep{lee68,brown94}, and then taking the median
value. The derived conversion factor, valid for the summed and MOS datasets,
is CF $=6.6 \times 10^{-12}$~erg~cm$^{-2}$~cnt$^{-1}$, with an uncertainty
of $\sim 15$\% ($1\sigma$ standard deviation). For the source detected only
in the PN we derived in the same way a median conversion factor CF $=2.1
\times 10^{-12}$~erg~cm$^{-2}$~cnt$^{-1}$ for the PN count rates. We applied
these factors to all stars for which no spectral analysis can be performed,
assuming that all of them have the mean absorption, since individual
absorption measures are generally not available. For stars having
$N_\mathrm{H} \sim 1-2 \times 10^{21}$ cm$^{-2}$, as found for some sources,
this will underestimate the X-ray luminosity by a factor of $\sim 1.5$. For
the stars with spectral fits (Table~\ref{tabfits} and Paper~I) we used
instead the fluxes determined from the PN or MOS1 (if PN is not available)
best-fit models. Finally, X-ray luminosities have been derived using the
{\it Hipparcos} cluster distance of 352 pc.

The sensitivity in the center of the field (from 3\,$\sigma$ upper limits)
is $L_{\rm X} \sim 2 \times 10^{28}$~erg~s$^{-1}$, and decreases to $\sim
4\times 10^{28}$~erg~s$^{-1}$ at 13$\arcmin$ offaxis, and to $\sim 7\times
10^{28}$~erg~s$^{-1}$ where only MOS is present.

Fig.~\ref{lx-col} shows $L_\mathrm{X}$ and $L_\mathrm{X}/L_\mathrm{bol}$ as
a function of the $R-I$ colour for cluster members and candidates. For
early-type stars, we find $\log L_\mathrm{X}/L_\mathrm{bol} \sim -6.6$ for
the O9.5 star $\sigma$~Ori~AB, consistent with the typical value ($\sim -7$)
found for hot stars \citep{pallavic81,bergh97}, while $\log
L_\mathrm{X}/L_\mathrm{bol} \sim -5.5$ for the other early-type stars and
for the F7 star. For late-type stars there is a general trend of decreasing
$L_\mathrm{X}$ with increasing colour, although there is a scatter of more
than one order of magnitude at each colour. Correspondingly,
$L_\mathrm{X}/L_\mathrm{bol}$ is nearly constant, as observed in other young
clusters and star-forming regions 
\citep[SFRs; e.g.][]{feigel02,flaccomio03o2,flaccomio03ev}. The median value
is $\log L_\mathrm{X}/L_\mathrm{bol} \sim -3.3$ for detected stars, and
$\log L_\mathrm{X}/L_\mathrm{bol}\sim -3.4$ taking into account also upper
limits, i.e. very close to the saturation limit. This value is consistent
with the results obtained for other SFRs of similar age
\citep[e.g.][]{flaccomio03ev}. There are a few stars lying significantly
above the saturation value $\log L_\mathrm{X}/L_\mathrm{bol}\sim -3$: this
is generally due to the presence of flares or to strong variability during
the observation.

\begin{figure}
\resizebox{\hsize}{!}{\includegraphics{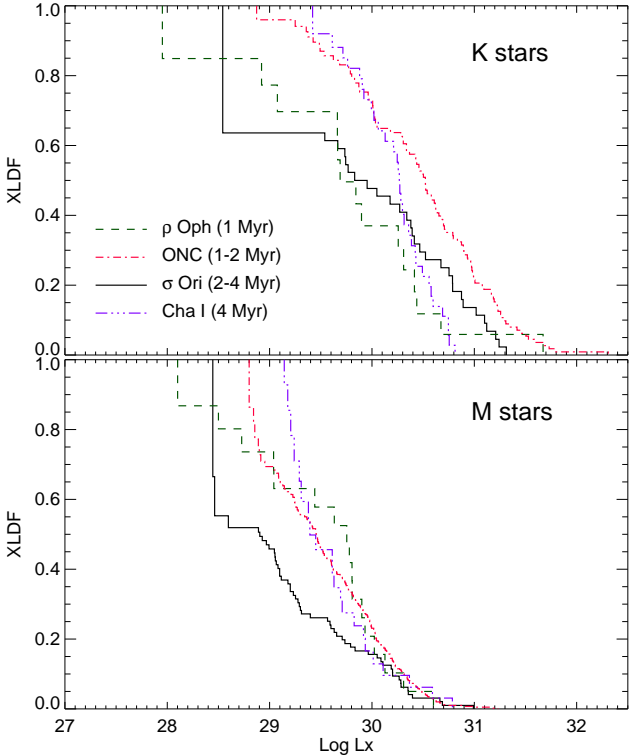}}
\caption{Comparison of the XLDF of $\sigma$~Ori with those of other SFRs and
associations of similar age for K and early M stars}
\label{xldf-sfr}
\end{figure}

In Fig.~\ref{xldf} we show the X-ray luminosity distribution function (XLDF)
for K and early-M stars (upper panel). The XLDF confirms the decrease of
X-ray luminosity with spectral type: K-type stars are clearly more luminous
than M-type stars, with a median $\log L_\mathrm{X} =29.73$ erg~s$^{-1}$ for
K stars, an order of magnitude higher than the median $\log L_\mathrm{X}
=28.77$ erg~s$^{-1}$ for M-type stars.

For the subsample of stars with available H$\alpha$ measurements we also
compared the XLDFs for classical (CTTS) and weak-lined (WTTS) T~Tauri stars.
We have classified stars as CTTS or WTTS using the H$\alpha$ EW criterium
derived by \citet{barrado03ha}, which depends on spectral type. In the
bottom panels of Fig.~\ref{xldf} we plot the XLDFs for CTTS and WTTS
separately for stars ealier than M5, and for stars and brown dwarfs of all
spectral types. In the first case, the two distributions are clearly
different, with a median $\log L_\mathrm{X} = 30.12$ erg~s$^{-1}$ for WTTS
and $\log L_\mathrm{X} = 29.47$ erg~s$^{-1}$ for CTTS. As a check, we have
performed a series of two-sample tests, as implemented in {\sc asurv} Rev.
1.1 \citep{asurv}, finding a probability $P=0.04-0.11$ that the two
distributions are drawn from the same population. If one considers all stars
and brown dwarfs, the two distributions do not appear to be significantly
different, although CTTS are still a factor of $\sim 2$ less luminous
(median $\log L_\mathrm{X} \sim 28.76$) than WTTS (median $\log
L_\mathrm{X}\sim 29.11$). In this latter case the two-sample tests are
inconclusive, giving $P=0.7-0.9$. The different results found in the two
cases can be attributed to the higher number of upper limits in the full
sample of stars and brown dwarfs: in fact, for spectral types later than
$\sim\,$M5 we detected only one CTTS out of 9 and none of the 19 WTTS,
compared with a detection rate of 18/24 ($\sim$\,75\%) and 24/27
($\sim$\,90\%) for CTTS and WTTS earlier than M5, respectively.

In Fig.~\ref{xldf-sfr} we compare the XLDFs of K and M-type stars of
$\sigma$~Ori with those of other SFRs and associations of similar age. We
have used for the comparison the data of \object{$\rho$~Oph} (1~Myr),
\object{Orion Nebula Cluster} (ONC, 1--2~Myr) and \object{Chameleon~I}
(4~Myr) reanalyzed by \citet[][see their paper for details]{flaccomio03ev}.
Since the luminosities derived by these authors are in the 0.1--4 keV band,
we have computed an appropriate conversion factor using the same procedure
described above, in order to derive luminosities in the same band. We find
that, for K-type stars, the high-luminosity tail ($\log L_\mathrm{X} >
30.5$) is intermediate between that of ONC and those of $\rho$~Oph and
Cha~I, and the median luminosity appears to be significantly lower than ONC
and Cha~I (by a factor of $\sim 3-5$), but comparable to that of $\rho$~Oph.
The two sample tests indicate that the XLDF is significantly different from
that of ONC ($P<0.002$) but are inconclusive in the other cases. On the
other hand, for M-type stars the high-luminosity tails of all the SFRs are
comparable, but the median for $\sigma$~Ori ($\log L_\mathrm{X} = 28.9$) is
significantly lower than those of all the other SFRs ($\log L_\mathrm{X} =
29.4-29.7$); the two-sample tests confirm that the XLDFs are not drawn by
the same population, giving in all cases $P< 0.0002$.

\begin{figure*}
\resizebox{\hsize}{!}{\includegraphics[clip]{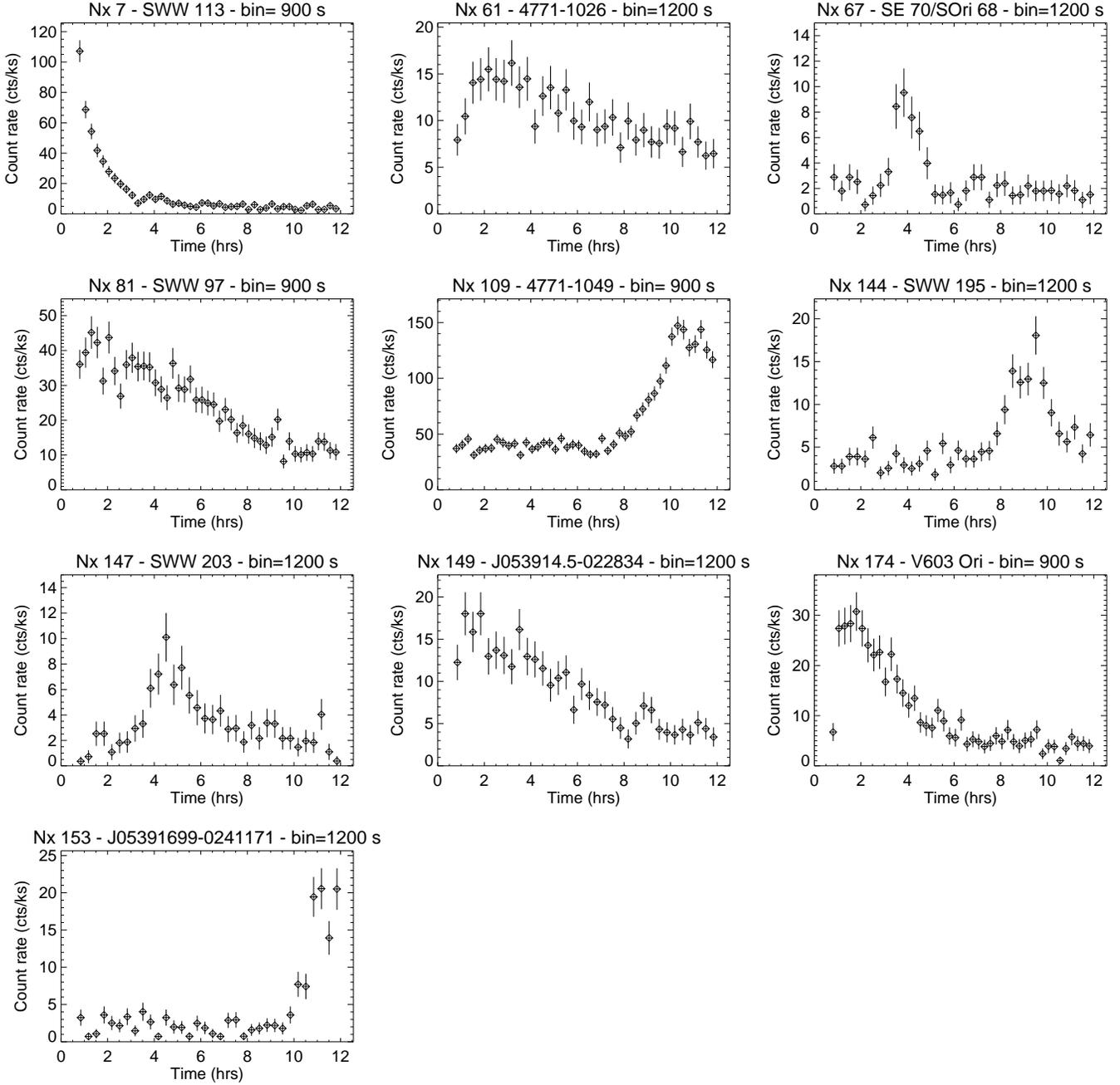}}
\caption{Combined PN+MOS1+MOS2 light curves of cluster members or candidates
showing strong flares during our observation, including a candidate member
from 2MASS (last panel). Count rates are espressed as MOS equivalent count
rates. The different bin size used for each source is indicated at the top
of each panel together with the source identification}
\label{flares}
\end{figure*}

\begin{figure*}
\resizebox{\hsize}{!}{\includegraphics[clip]{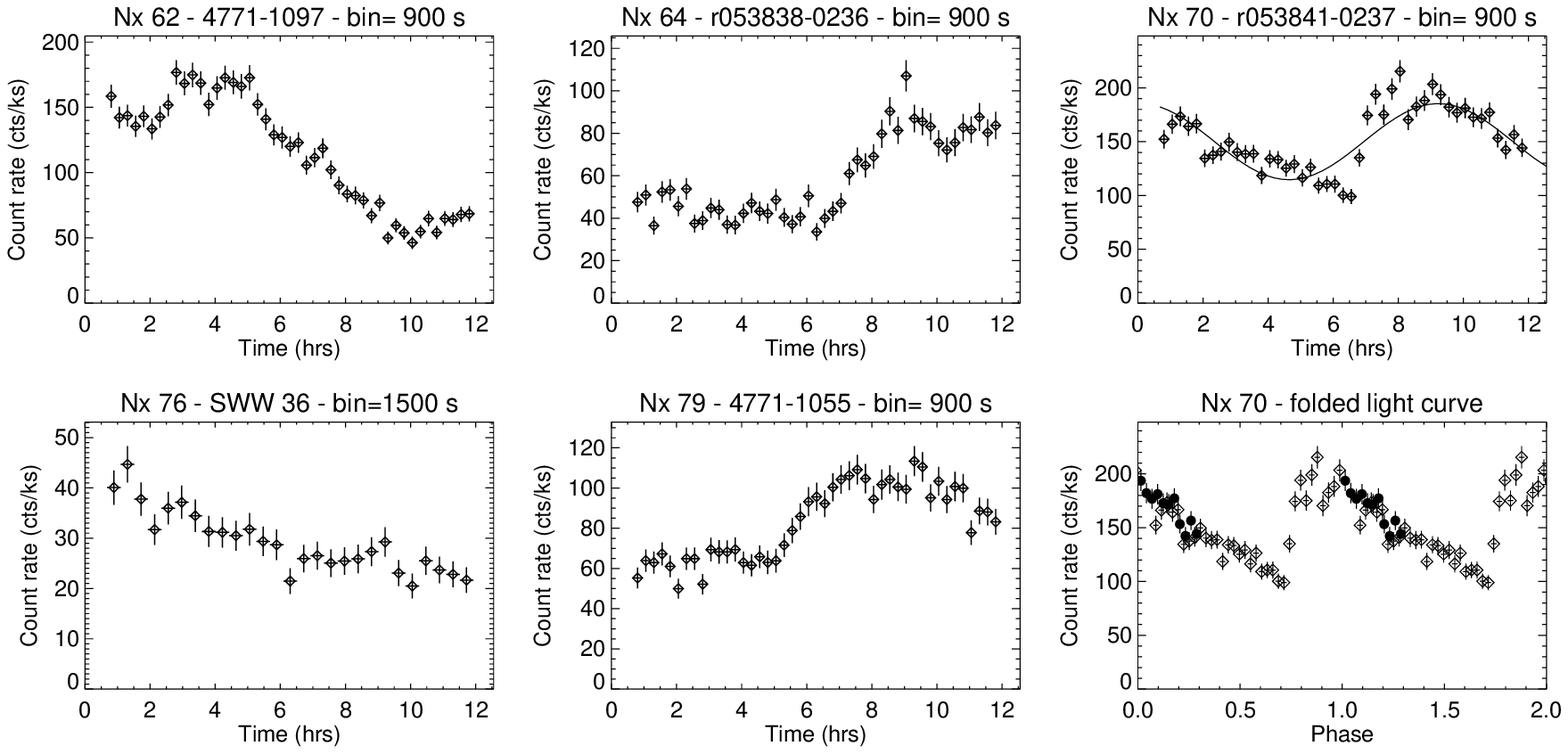}}
\caption{Combined PN+MOS1+MOS2 light curves of cluster members or candidates
showing significant variability not clearly attributable to flares. Note in
particular the modulation of source NX~70 ({\em right panels}). The top
right panel shows the light curve of NX~70 together with a best-fit
sinusoidal curve (solid line). In the bottom right panel we show the same
light curve folded with the derived rotational period of 9.2 hours, where
different symbols mark data from the first ({\em diamonds}) and second ({\em
filled circles}) periods; for clarity two rotational phases are shown
}
\label{slowvar}
\end{figure*}

\subsection{Variability}
\label{var}

Light curves for all detected members or candidates have been extracted from
the photon event lists using circular regions of radius $24\arcsec$ centered
on the source position, except for a few cases where we used smaller radii
to avoid overlap with nearby close sources. Inspection of the light curves
reveals that many of the detected members or candidates show significant
variability during our observation, including the strong X-ray flare
detected on the B2Vp star $\sigma$~Ori~E and discussed in Paper~I. In order
to quantify the variability of the sources, we have performed a
Kolmogorov-Smirnov test on the unbinned photon arrival times, using data
from PN, or from MOS for sources falling on PN CCD gaps or outside of the PN
field of view. We find that $\sim 35$\% (29/84) of late-type members are
variable at the 99\% confidence level, and other 7 stars (8\%) are possibly
variable at the 95\% level. In particular, 47\% of the CTTS and 54\% of the
WTTS are variable at the 99\% level. In addition, variability at the 99\%
level is detected in two of the 2MASS candidates of Table~\ref{detcand}.
Ten late-type sources show clear flares, including one of the 2MASS
candidates, with increases in the count rates by factors of $\sim 4-20$ in
$\sim 1-2$ hours and decay times of $\sim 2-10$ hours. The combined
PN+MOS1+MOS2 light curves of these flaring sources (espressed as MOS
equivalent count rates) are shown in Fig.~\ref{flares}. The other variable
sources show generally low-level flare-like variability or slow variations.
In Fig.~\ref{slowvar} we show some examples of light curves of sources with
strong variability not clearly attributable to flares. One source (NX~76)
shows a steady decay by a factor of $\sim 2$ during the entire observation,
that could be attributed either to the decay of a long-lasting flare or to
rotational modulation. Sources NX~62 and NX~64 have very high temperatures
(see Table~\ref{tabfits}), but their light curve does not show the typical
behaviour of flares: in both cases a steady emission level for a few hours
is observed after a slow rise or before a slow decay. A similar behaviour is
observed for NX~79. All these light curves could be interpreted as due to
rotational modulation of the emission from active, possibly flaring regions
unevenly distributed on the stellar surface \citep{stelzer99}. A more
evident modulation effect is instead present in source NX~70 (source \#4 in
Paper~I), which shows two peaks followed by a similar decay trend. A fit of
the light curve with a simple sinusoidal model gives a period of $\sim 9.2
\pm 0.1$ hours, with an amplitude variation of $\sim 25$\% around the mean.
The folded light curve is shown in the bottom right panel of
Fig.~\ref{slowvar}, where we have marked with different symbols data points
belonging to different periods: there is a perfect overlap between the light
curves in the two periods. We caution, however, that, having data available
only for slightly more than one period, these results are not conclusive,
and we cannot completely exclude that the observed variability is due
to the occurrence of two similar flaring events rather than to
rotational modulation.

\section{Spectral analysis}
\label{spectra}

For the detected cluster members or candidates having at least 500
cts in the PN or MOS1 we performed spectral analysis. We excluded sources
falling on CCD gaps, that have an effective low number of counts despite
their derived high count rate from the wavelet detection. A detailed PN and
MOS spectral analysis of the four bright central sources has already been
performed in Paper~I; however, we have reanalyzed the spectra of the two
central K-type stars (NX\,65 $=$ source \#3 of Paper~I, and NX\,70 $=$
source \#4) in order to compare in a consistent way their properties with
those of the other late-type sources discussed in this paper.

PN and/or MOS1 spectra have been extracted from the same circular regions
used for the light curves. Background spectra for each source were
extracted from a nearby circular region free from other X-ray sources and on
the same CCD chip, using the same extraction radius as the corresponding
source region. Response matrices and ancillary files were generated for each
source using the standard SAS tasks {\sc rmfgen}  and {\sc arfgen}. Spectra
have been rebinned to have at least 20 counts per bin, and were
fitted in XSPEC v.11.3.0 in the energy range $0.3-8$ keV, using a
two-temperature APEC v.1.3.0 model with variable global abundance. For most
of the weakest sources with $<1000$ cts, acceptable fits were found with
only one temperature component. The hydrogen column density was generally
kept fixed at the value $N_\mathrm{H} = 2.7 \times 10^{20}$ cm$^{-2}$,
derived from the measured reddening $E(B-V)=0.05$. In a few cases however
this led to an unacceptable fit at low energies, and therefore we left the
column density as a free parameter, obtaining best-fit values of
$N_\mathrm{H} \sim 1.3-2 \times 10^{21}$ cm$^{-2}$ for quiescent sources,
i.e. higher by about one order of magnitude, and even higher values
($N_\mathrm{H} \ge 5\times 10^{21}$ cm$^{-2}$) for three flaring sources. We
note that the eight sources needing a higher column density are either CTTS
or stars with known IR excess; the only exception is source NX~145,
identified with the star \object{4771-1038}, a WTTS with no IR excess
\citep{oliveira04}.

\begin{table*}
\caption{Results of the 2-T or 1-T spectral fitting of the PN and/or MOS1
spectra of sources with more than 500 counts. Errors are 90\% confidence ranges
for one interesting parameter}
\label{tabfits}
\begin{tabular}{rcccccclrcll}
\hline\hline
\noalign{\smallskip}
NX& $T_1$& $T_2$ & $EM_1$ & $EM_2$& $Z/Z_\odot$& 
  $N_\mathrm{H}^{\,a}$& $\chi^2_r$/dof& $F_\mathrm{X}^{\,b}$ & Inst.$^{\,c}$& 
  CTT/& Notes\\
& \multicolumn{2}{c}{(keV)}&\multicolumn{2}{c}{($10^{53}$ cm$^{-3}$)}& 
  & & & & & WTT& \\
\hline
\noalign{\smallskip}
  2& $0.75_{-0.13}^{+0.09}$& $1.63_{-0.26}^{+0.44}$ &  $5.19_{-2.55}^{+5.84}$& 
     $8.90_{-3.20}^{+4.09}$& $0.25_{-0.10}^{+0.17}$&                       & 
     0.56/59 &  9.10& M& CTT?&                \\[2pt]
  3& $0.83_{-0.06}^{+0.15}$& $2.39_{-0.61}^{+1.03}$ & $12.83_{-2.88}^{+2.71}$& 
     $6.64_{-1.47}^{+1.73}$& $0.16_{-0.05}^{+0.07}$&                       & 
     0.62/74 & 11.55& M& WTT?&                \\[2pt]
  7& $4.83_{-1.42}^{+2.86}$&                        &  $3.75_{-0.76}^{+1.01}$& 
                           & $0.56_{-0.49}^{+0.72}$& $5.79_{-1.32}^{+1.89}$& 
     0.67/34 &  3.99& P& CTT & flare          \\[2pt]
  8& $0.61_{-0.22}^{+0.14}$& $1.25_{-0.16}^{+0.19}$ &  $1.69_{-0.56}^{+1.10}$& 
     $3.40_{-1.30}^{+1.47}$& $0.23_{-0.08}^{+0.13}$&                       & 
     0.63/80 &  3.05& P&     &                \\[2pt]
   & $0.43_{-0.14}^{+0.36}$& $1.02_{-0.50}^{+0.31}$ &  $2.11_{-0.98}^{+1.45}$& 
     $3.65_{-1.35}^{+1.68}$& $0.14_{-0.06}^{+0.06}$&                       & 
     0.97/24 &  2.64& M&     &          \\[2pt]
 34& $0.80_{-0.14}^{+0.20}$&                        &  $1.88_{-0.40}^{+0.50}$& 
                           & $0.02_{-0.02}^{+0.04}$&                       & 
     0.51/21 &  0.61& P&     &                \\[2pt]
 39& $0.56_{-0.25}^{+0.22}$& $1.26_{-0.25}^{+0.37}$ &  $0.56_{-0.25}^{+0.40}$& 
     $1.14_{-0.30}^{+0.40}$& $0.12_{-0.06}^{+0.13}$&                       & 
     0.59/39 &  0.82& P&     &                \\[2pt]
 49& $0.29_{-0.05}^{+0.07}$& $1.01_{-0.10}^{+0.19}$ &  $1.11_{-0.44}^{+0.64}$& 
     $1.54_{-0.47}^{+0.58}$& $0.19_{-0.07}^{+0.13}$&                       & 
     0.64/60 &  1.24& P&     &                \\[2pt]
 61& $1.92_{-0.49}^{+0.72}$&                        &  $2.28_{-0.93}^{+1.71}$& 
                           & $0.29_{-0.26}^{+0.59}$& $5.00_{-1.52}^{+2.76}$& 
     0.47/19 &  1.60& P& CTT & IR exc.,flare \\[2pt]
 62& $1.06_{-0.11}^{+0.17}$& $7.69_{-2.63}^{+6.43}$ &  $9.76_{-0.87}^{+2.00}$& 
     $3.90_{-0.55}^{+0.62}$& $0.05_{-0.02}^{+0.03}$&                       & 
     0.50/80 &  8.14& M& WTT &                \\[2pt]
 64& $0.80_{-0.05}^{+0.06}$& $6.59_{-2.48}^{+8.26}$ &  $4.18_{-0.55}^{+0.56}$& 
     $1.13_{-0.19}^{+0.24}$& $0.08_{-0.03}^{+0.04}$&                       & 
     0.57/131&  2.82& P& WTT & no IR exc.     \\[2pt]
   & $0.80_{-0.13}^{+0.19}$& 6.59$^{\,d}$           &  $3.91_{-1.14}^{+1.16}$& 
     $1.04_{-0.43}^{+0.33}$& $0.07_{-0.04}^{+0.05}$&                       & 
     0.51/40 &  2.53& M&     &                \\[2pt]
 65& $0.94_{-0.04}^{+0.04}$& $3.65_{-0.55}^{+0.67}$ &  $8.84_{-0.68}^{+0.68}$& 
     $4.34_{-0.49}^{+0.43}$& $0.12_{-0.03}^{+0.03}$&                       &
     0.69/298&  7.89& P& WTT &                \\[2pt]
   & $0.97_{-0.24}^{+0.11}$& $3.63_{-0.86}^{+1.58}$ & $13.97_{-9.09}^{+4.12}$& 
     $4.48_{-0.80}^{+8.17}$& $0.06_{-0.02}^{+0.03}$&                       &
     0.74/111&  9.49& M&     &                \\[2pt]
 69& $0.30_{-0.05}^{+0.05}$& $1.22_{-0.28}^{+0.15}$ &  $3.88_{-1.10}^{+0.90}$& 
     $2.61_{-0.65}^{+3.04}$& $0.11_{-0.07}^{+0.07}$& $1.88_{-0.37}^{+0.32}$& 
     0.62/43 &  2.22& P& CTT & IR exc.        \\[2pt]
 70& $0.76_{-0.04}^{+0.04}$& $2.54_{-0.20}^{+0.26}$ &  $1.93_{-0.46}^{+0.70}$&
     $6.32_{-0.42}^{+0.43}$& $0.35_{-0.09}^{+0.11}$&                       &
     0.81/283&  6.60& P& CTT &                \\[2pt]
   & $0.80_{-0.08}^{+0.24}$& $3.01_{-0.54}^{+0.71}$ &  $4.08_{-2.14}^{+4.24}$&
     $6.36_{-0.73}^{+0.85}$& $0.17_{-0.10}^{+0.20}$&                       &
     0.68/95 &  7.25& M&     &                \\[2pt]
 76& $0.78_{-0.09}^{+0.07}$& $2.46_{-0.89}^{+1.75}$ &  $1.19_{-0.61}^{+0.99}$& 
     $0.79_{-0.21}^{+0.28}$& $0.21_{-0.10}^{+0.22}$&                       & 
     0.70/70 &  1.29& P&     &                \\[2pt]
   & $0.71_{-0.13}^{+0.13}$& $1.37_{-0.41}^{+0.70}$ &  $1.10_{-0.59}^{+0.86}$& 
     $1.13_{-0.42}^{+0.87}$& $0.20_{-0.11}^{+0.20}$&                       & 
     0.42/20 &  1.29& M&     &          \\[2pt]
 78& $0.33_{-0.06}^{+0.08}$& $1.29_{-0.10}^{+0.06}$ &  $1.94_{-0.52}^{+0.55}$& 
     $4.91_{-0.65}^{+0.79}$& $0.15_{-0.05}^{+0.06}$&                       & 
     0.67/133&  3.23& P& WTT &                \\[2pt]
   & $0.33_{-0.11}^{+0.51}$& $1.33_{-0.15}^{+0.29}$ &  $1.32_{-0.67}^{+0.87}$& 
     $4.67_{-1.41}^{+1.44}$& $0.17_{-0.09}^{+0.23}$&                       & 
     0.89/40 &  3.05& M&     &          \\[2pt]
 79& $0.79_{-0.07}^{+0.05}$& $1.94_{-0.28}^{+0.46}$ &  $2.61_{-0.80}^{+1.42}$& 
     $4.26_{-0.86}^{+0.59}$& $0.20_{-0.06}^{+0.08}$&                       & 
     0.72/177&  4.32& P&     &                \\[2pt]
   & $0.77_{-0.21}^{+0.29}$& $1.84_{-0.49}^{+0.82}$ &  $3.51_{-1.45}^{+1.44}$& 
     $4.33_{-0.85}^{+1.10}$& $0.09_{-0.06}^{+0.10}$&                       & 
     0.88/52 &  4.05& M&     &          \\[2pt]
 81& $0.86_{-0.11}^{+0.15}$& $3.21_{-0.61}^{+0.89}$ &  $0.18_{-0.09}^{+0.25}$& 
     $1.28_{-0.30}^{+0.27}$& $0.87_{-0.47}^{+0.86}$&                       & 
     0.62/57 &  1.60& P&     & flare          \\[2pt]
 90& $0.97_{-0.13}^{+0.07}$& $3.31_{-0.77}^{+0.64}$ &  $8.17_{-3.60}^{+4.82}$& 
     $8.20_{-3.66}^{+1.68}$& $0.14_{-0.08}^{+0.19}$& $1.64_{-0.33}^{+0.45}$& 
     0.56/181& 10.91& P& WTT & IR exc.        \\[2pt]
   & $0.93_{-0.20}^{+0.15}$& $3.35_{-0.77}^{+1.10}$ &  $9.15_{-5.34}^{+8.82}$& 
     $9.45_{-1.48}^{+1.93}$& $0.12_{-0.08}^{+0.33}$& $1.91_{-0.55}^{+0.74}$& 
     0.73/66 & 12.07& M&     &          \\[2pt]
 92& $0.32_{-0.04}^{+0.07}$& $1.49_{-0.31}^{+0.23}$ &  $0.52_{-0.25}^{+0.43}$& 
     $0.93_{-0.36}^{+0.52}$& $0.47_{-0.30}^{+0.65}$&                       & 
     0.70/54 &  1.07& P&     &                \\[2pt]
 93& $0.63_{-0.13}^{+0.14}$& $1.27_{-0.18}^{+0.25}$ &  $0.76_{-0.27}^{+0.53}$& 
     $1.17_{-0.61}^{+0.74}$& $0.32_{-0.11}^{+0.12}$&                       & 
     0.51/74 &  1.36& P& CTT & no IR exc.     \\[2pt]
   & $0.85_{-0.08}^{+0.15}$&                        &  $3.37_{-1.22}^{+1.11}$& 
                           & $0.08_{-0.03}^{+0.05}$&                       & 
     0.45/22 &  1.41& M&     &          \\[2pt]
104& $0.37_{-0.16}^{+0.32}$& $0.84_{-0.12}^{+0.21}$ &  $1.16_{-0.62}^{+0.80}$& 
     $1.78_{-0.76}^{+0.71}$& $0.09_{-0.05}^{+0.11}$&                       & 
     0.37/28 &  1.04& P& WTT &                \\[2pt]
106& $0.61_{-0.26}^{+0.13}$& $1.43_{-0.32}^{+0.55}$ &  $0.47_{-0.24}^{+0.39}$& 
     $0.74_{-0.24}^{+0.39}$& $0.35_{-0.18}^{+0.41}$&                       & 
     0.56/31 &  0.89& P& WTT &                \\[2pt]
109& $0.98_{-0.15}^{+0.13}$& $6.84_{-2.78}^{+10.68}$&  $5.84_{-1.08}^{+1.17}$& 
     $2.08_{-0.33}^{+0.37}$& $0.10_{-0.04}^{+0.20}$&                       & 
     0.74/58 &  4.86& M& WTT & flare          \\[2pt]
110& $0.93_{-0.13}^{+0.12}$&                        &  $7.07_{-2.37}^{+2.85}$& 
                           & $0.09_{-0.05}^{+0.07}$&                       & 
     0.68/20 &  3.12& M& WTT &                \\[2pt]
122& $0.65_{-0.24}^{+0.21}$& $1.27_{-0.26}^{+0.55}$ &  $0.24_{-0.13}^{+0.43}$& 
     $0.58_{-0.25}^{+0.43}$& $0.38_{-0.21}^{+0.51}$&                       & 
     0.44/26 &  0.62& P& WTT?&                \\[2pt]
132& $0.73_{-0.32}^{+0.22}$& $1.67_{-0.56}^{+1.55}$ &  $0.58_{-0.36}^{+1.25}$& 
     $0.74_{-0.27}^{+0.47}$& $0.24_{-0.17}^{+0.34}$&                       & 
     0.40/30 &  0.84& P& WTT & no IR exc.     \\[2pt]
138& $0.84_{-0.16}^{+0.27}$& $2.86_{-0.69}^{+4.92}$ &  $0.77_{-0.52}^{+2.52}$& 
     $2.02_{-1.16}^{+0.55}$& $0.26_{-0.22}^{+0.64}$& $2.08_{-0.72}^{+1.43}$& 
     0.68/42 &  2.14& P& CTT?&                \\[2pt]
145& $0.23_{-0.04}^{+0.09}$& $1.06_{-0.13}^{+0.18}$ &  $2.19_{-1.26}^{+2.18}$& 
     $1.36_{-0.19}^{+0.46}$& $0.19_{-0.10}^{+0.43}$& $1.32_{-0.99}^{+2.21}$& 
     0.44/36 &  1.40& P& WTT & no IR exc.     \\[2pt]
149& $1.21_{-0.40}^{+0.72}$&                        &  $2.08_{-0.85}^{+2.08}$& 
                           & $0.08_{-0.07}^{+0.12}$& $1.48_{-0.72}^{+1.11}$& 
     0.51/19 &  0.99& P& WTT & IR exc.?,flare\\[2pt]
150& $0.85_{-0.09}^{+0.16}$&                        &  $1.29_{-0.33}^{+0.34}$& 
                           & $0.11_{-0.05}^{+0.08}$&                       & 
     0.69/22 &  0.59& P&     & A8V          \\[2pt]
156& $0.56_{-0.16}^{+0.14}$& $1.23_{-0.22}^{+0.85}$ &  $0.80_{-0.25}^{+0.55}$& 
     $1.14_{-0.42}^{+0.44}$& $0.30_{-0.13}^{+0.23}$&                       & 
     0.81/38 &  1.29& P&     &                \\[2pt]
170& $0.36_{-0.09}^{+0.37}$& $1.24_{-0.27}^{+0.32}$ &  $0.64_{-0.33}^{+0.49}$& 
     $0.99_{-0.44}^{+0.70}$& $0.33_{-0.21}^{+0.69}$&                       & 
     0.34/24 &  1.03& P&     &                \\[2pt]
172& $0.74_{-0.08}^{+0.05}$& $2.02_{-0.30}^{+0.54}$ &  $2.03_{-0.65}^{+0.93}$& 
     $4.73_{-0.76}^{+0.76}$& $0.54_{-0.18}^{+0.33}$&                       & 
     0.85/105&  6.07& P& WTT & G5, no IR exc.     \\[2pt]
174& $5.44_{-1.54}^{+5.58}$&                        &  $7.47_{-1.45}^{+1.59}$& 
                           & $0.74_{-0.64}^{+0.85}$&  $28.5_{-8.0}^{+9.0}$ & 
     0.54/24 &  8.38& P& CTT & IR exc.,flare  \\
\noalign{\smallskip}
\hline
\noalign{\smallskip}
\multicolumn{12}{l}{$a$: hydrogen column density in units of $10^{21}$
cm$^{-2}$. Where no value is given, it was kept fixed at $N_{\rm H} = 2.7 \times
10^{20}$ cm$^{-2}$}\\
\multicolumn{12}{l}{$b$: unabsorbed X-ray flux in the $0.3-8$ keV band, in units
of $10^{-13}$ erg\,cm$^{-2}$\,s$^{-1}$}\\
\multicolumn{12}{l}{$c$: instrument from which the spectrum was extracted (P for
PN, M for MOS1)}\\
\multicolumn{12}{l}{$d$: $T_2$ was kept fixed to the PN fit value since it was
not constrained by the MOS1 fit}\\
\end{tabular}
\end{table*}

\begin{figure*}
\resizebox{\hsize}{!}{\includegraphics[clip]{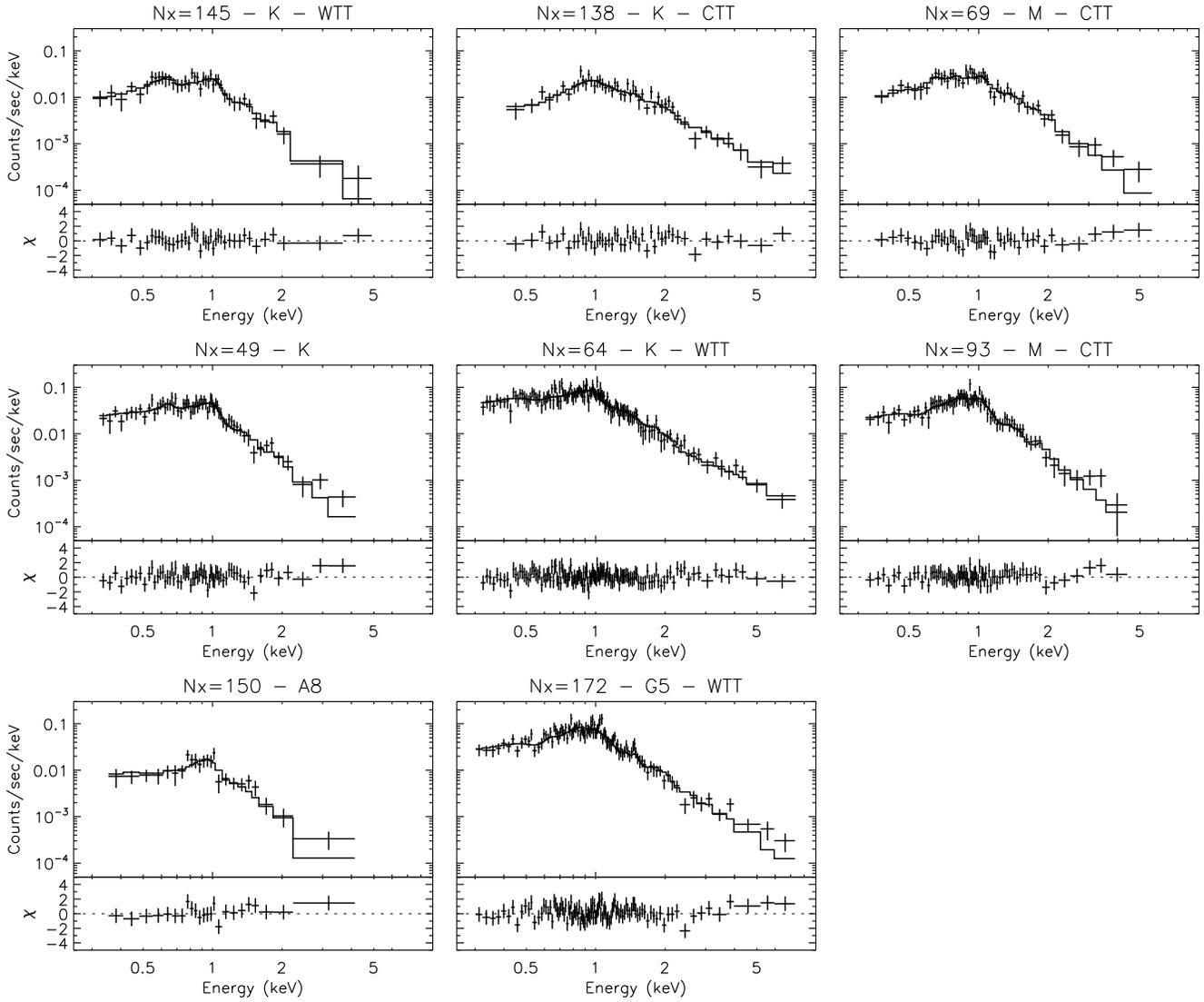}}
\caption{{\em Top and middle panels}: examples of PN spectra of K-type and
M-type cluster members with high ($N_\mathrm{H}\sim 1-2\times 10^{21}$
cm$^{-2}$, {\em top panels}) and low ($N_\mathrm{H}= 2.7\times 10^{20}$
cm$^{-2}$, {\em middle panels}) absorption. For each $N_\mathrm{H}$ case,
two spectra are shown for K-type stars, one corresponding to the coolest
($<T>\sim 0.6$ keV, {\em left}) and the other to the hottest ($<T> \sim 2$
keV, {\em center}) coronal temperatures; for M-type stars we show in both
cases typical spectra with $<T>\sim 1$ keV ({\em right}).
{\em Bottom panels}: PN spectra of the A8 and the G5 stars. Note the
similarity of both spectra with those of later type stars}
\label{spq}
\end{figure*}

\begin{figure*}
\resizebox{\hsize}{!}{\includegraphics[clip]{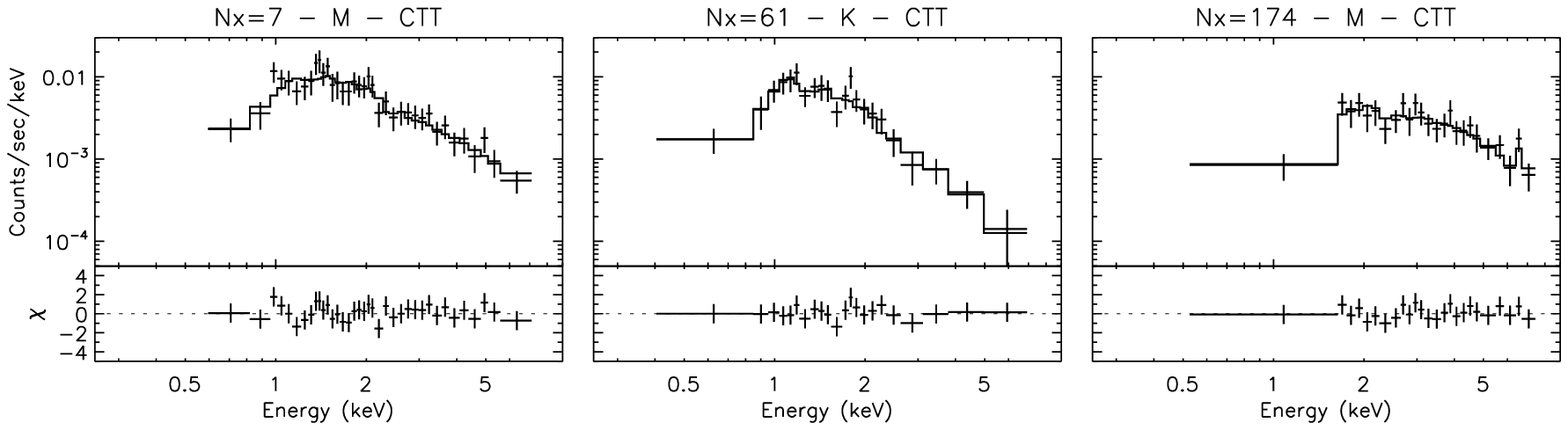}}
\caption{PN spectra of the three CTT stars showing strong flares with high
absorption ($N_\mathrm{H} \ge 5\times 10^{21}$ cm$^{-2}$). Note in
particular, for source NX~174, the very high absorption that strongly
suppresses all emission below 1.5 keV, and the evident Fe 6.7 keV line}
\label{spfl}
\end{figure*}

The resulting best-fit parameters are given in Table~\ref{tabfits}, and a
sample of the spectra together with the best-fit models are shown in
Fig.~\ref{spq} for quiescent sources (with high and low absorption) and in
Fig.~\ref{spfl} for the three flaring CTT stars with higher absorption.
We find typical quiescent temperatures of $T_1\sim 0.3-0.8$ keV and $T_2\sim
1-3$ keV, in agreement with the results found for other young clusters and
SFRs \citep[e.g][]{feigel99}. In two cases (NX~62 and NX~64) temperatures up
to $6-8$ keV are found, although it is not clear from their
light curves whether these sources are flaring during the observation (see
Fig.~\ref{slowvar}). Emission measures range between $10^{52}$ and $10^{54}$
cm$^{-3}$, with $EM_2/EM_1 \sim 0.5-2.5$. In all cases we find subsolar
abundances, with most stars having $Z\sim 0.1-0.3\,Z_\odot$, as commonly
observed in active stars. Somewhat higher abundances ($Z\sim
0.6-0.9\,Z_\odot$) are found for three flaring sources. For the small
subsample of WTTS and CTTS we find no significant difference in the coronal
parameters of the two classes: in particular we do not find the higher
spread in CTTS abundances with respect to WTTS found in \object{L1551}
by \citet{favata03sfr}.

Particularly interesting is source NX~174, identified with
\object{V603~Ori}, a Class~II star with strong IR excess and source of the
protostellar jet \object{HH\,445} \citep{olivloon04}. This source underwent
a strong flare during the observation: the spectrum is quite hot, with
$T\sim 7$ keV, showing an evident Fe 6.7 keV line, and the emission is
strongly suppressed below 1.5 keV, implying a column density $N_\mathrm{H} =
2.8\times 10^{22}$ cm$^{-2}$, i.e. two orders of magnitude higher than the
mean cluster absorption.

\section{Discussion and conclusions}
\label{concl}

In this paper we have presented the spatial analysis of an XMM-{\it Newton}
observation of the $\sigma$~Ori cluster centered on the hot star
$\sigma$~Ori~AB. We have detected half of the early-type (O-A) stars,
including $\sigma$~Ori~AB, 65\% of F-K stars and $\sim\,$50\% of early M
(i.e. M0--M4) stars above a detection limit of $\sim 2\times 10^{28}$
erg~s$^{-1}$. Our detection rates for late-type stars are lower than what
observed in other SFRs and very young clusters, where $\sim 80$\% of
late-type stars are generally detected above similar levels
\citep[e.g.][]{garmire00,preibisch01,flaccomio03o1,flaccomio03ev}. A
possible explanation for this discrepancy might be contamination of our
optical sample by non-members. While other regions have well
established membership lists, in the case of $\sigma$~Ori only $\sim 50$\%
of the candidate and probable members have been confirmed spectroscopically.
On the other hand, it is well known that PMS stars are powerful X-ray
emitters, with luminosities $10-10^4$ times higher than older late-type
stars \citep{feigel99}, and the presence of strong X-ray emission is
generally used to identify new members of SFRs and young clusters. It is
therefore likely that most of the undetected candidates might turn out to be
older field stars rather than young cluster members. In particular, there
are 45 undetected cluster candidates that have been selected on the basis of
photometry only: if most, or all, of them were indeed non-members, we would
raise the detection rate of late-type stars to $\sim 75-80$\%, bringing our
results in agreement with other observations. Optical spectroscopic
observations are needed (and in fact have been recently obtained by us using
the multi-object spectrograph FLAMES at the VLT; data analysis is in
progress) in order to determine the membership status of these stars and to
check whether the lower detection rate found in $\sigma$~Ori is due to the
inclusion of non-members in the sample, or it reflects a true difference
from other SFRs.

We have found that for late-type stars $L_\mathrm{X}$ declines towards later
spectral types, while $L_\mathrm{X}/L_\mathrm{bol}$ is nearly constant, as
commonly observed in other young clusters and SFRs. The spread of $1.5-2$
orders of magnitude observed at all colours in both $L_\mathrm{X}$ and
$L_\mathrm{X}/L_\mathrm{bol}$ is consistent with earlier results
\citep[e.g.][]{feigel99,preibisch02}. The median $\log
L_\mathrm{X}/L_\mathrm{bol}$ is $\sim -3.3$, i.e. very close to the
saturation limit. This value is higher than the value of $\sim -4$ generally
found for young ($\la 2$ Myr) SFRs
\citep{preibisch01,feigel02,feigel03,skinner03,flaccomio03ev}, but is
consistent with the observed increase of $L_\mathrm{X}/L_\mathrm{bol}$ with
age for stars with $M<M_\odot$, reaching the saturation value of $10^{-3}$
at $\sim 4$ Myr \citep{flaccomio03ev,stelzer04cha}. 

The comparison of the XLDFs of $\sigma$~Ori late-type stars with those of
other SFRs of similar age shows that the median X-ray luminosity of
$\sigma$~Ori late-type stars is significantly lower than the other SFRs. Our
results appear to be in disagreement with \citet{flaccomio03ev}, who found
no evidence for a significant evolution of the median X-ray luminosity of
pre-main sequence stars with $M=0.25-2 \,M_\odot$ up to $\sim 10$ Myr.
However, this disagreement could be significantly reduced, or removed, if we
consider the higher percentage of upper limits in our sample and the likely
contamination by non-members, as discussed above. For example, if we exclude
all undetected candidate members with only photometry available, the XLDF of
$\sigma$~Ori becomes similar to the other ones, in agreement with the lack
of PMS evolution found by \citet{flaccomio03ev}. 

About 40\% of the detected members show significant variability, with $\sim
10$\% undergoing strong flares during our observation. Strong variability is
commonly observed in up to $\sim\,$50\% of PMS stars in SFRs and very young
clusters. The frequency of flares is comparable to what found e.g. in
\object{IC~348} \citep{preibisch02}, \object{NGC~1333} \citep{getman02} and
\object{NGC~2264} \citep{ramirez04}, but it is much lower than in
$\rho$~Oph, where $\sim 35$\% of T~Tauri stars were found to be flaring by
XMM-{\it Newton} in an observation of comparable duration \citep{ozawa05}.
In addition to typical stellar flares, with a rapid rise followed by a slow
decay, some sources show slower systematic variations, with some evidence
for periodic variability, as observed also in other SFRs
\citep{preibisch02,skinner03}: such variations have been generally
attributed to the effect of rotational modulation of active regions coming
in and out of view as the star rotates \citep[see, e.g.][]{stelzer99}.

Spectral analysis of the brightest sources indicates that the coronae of
late-type stars in the $\sigma$~Ori cluster have typical temperatures of
$0.3-0.8$ keV and $1-3$ keV, with emission measure ratios of $EM_2/EM_1\sim
0.5-2.5$. In a few cases temperatures up to 7 keV have been found even in
the absence of clear flares. Our results are in very good agreement with the
coronal temperatures generally found in SFRs
\citep{feigel99,imanishi02,favata03sfr,preibisch04,stelzer04cha}. In
addition, we find for most stars a strongly subsolar metallicity ($Z\sim
0.1-0.3 \,Z_\odot$), consistent with the results found in other SFRs
\citep{imanishi02,favata03sfr,ozawa05}, and in active late-type coronae.

A long debated question in the study of PMS stars is whether there is a
difference between the X-ray properties of CTTS and WTTS, i.e. whether X-ray
emission is influenced by the presence of an accretion disk. Observations of
SFRs have shown no significant difference in the X-ray luminosity of CTTS
and WTTS when using as a discriminant the presence of infrared excess,
indicative of a circumstellar disk
\citep{preibisch01,getman02,feigel02,feigel03,gagne04}, while CTTS appear to
be significantly less luminous than WTTS when using accretion indicators
such as H$\alpha$ or CaII emission
\citep{stelzer01,preibisch02,flaccomio03,feigel04,stassun04}. The latter
results are confirmed by our observations: using a selection in terms of
H$\alpha$ emission, we find in fact that the X-ray emission from CTTS is
significantly weaker than WTTS. 

For the small subsample of bright WTTS and CTTS for which we were able to
perform spectral analysis, we find no significant difference in the
temperatures and metallicity of the two classes. Both CTTS and WTTS show
temperatures up to $\sim 3$ keV, and metallicities $Z\sim 0.1-0.5 Z_\odot$.
In L1551 \citet{favata03sfr} found similar temperatures but a higher
metallicity spread in CTTS with respect to WTTS, which is not confirmed by
our data. We stress, however, that both their sample and ours are very
small, preventing us to draw definite conclusions on possible abundance
differences between the two classes. The high temperatures we find for CTTS
are in constrast with the results obtained from high-resolution observations
of \object{TW~Hya}, whose corona shows a very low temperature ($T\sim 3$ MK)
and high density ($n_\mathrm{e} \ga 10^{12}$~cm$^{-3}$) that have been
interpreted as due to X-ray emission from an accretion shock
\citep{kastner02,stelzer04}; however, other high-resolution observations of
CTTS have shown the presence of hotter plasma
\citep{schmitt05,argi05,pallavic05asr}, similar to what found in WTTS
\citep{kastner04,argi04,scelsi05}, suggesting a magnetic origin for X-ray
emission. 

Finally, we clearly detected 5 very low-mass objects and brown dwarf
candidates down to M7.5, with the possible detection of other two objects
with ambiguous identification. X-ray luminosities range between $3\times
10^{28}$ and $6\times 10^{29}$ erg~s$^{-1}$, with
$L_\mathrm{X}/L_\mathrm{bol} \sim 10^{-3.6} - 10^{-2.5}$. Two of the
candidate brown dwarfs (\object{S\,Ori~3} and S\,Ori~25) show significant
variability, while a flare is detected in \object{SWW~203} ($=$ NX\,147), a
very-low mass star of spectral type $\sim$\,M5 (Fig.~\ref{flares}). Several
brown dwarfs and very-low mass stars have been detected in SFRs by {\it
ROSAT} \citep{neuh99,mokler02}, {\it Chandra}
\citep{imanishi01bd,preibisch01,feigel02,gagne04} and XMM-{\it Newton}
\citep{stelzer04cha,ozawa05}, with $L_\mathrm{X} \sim 10^{27}-10^{30}$
erg~s$^{-1}$ and $L_\mathrm{X}/L_\mathrm{bol} \sim 10^{-4.7}-10^{-2.6}$.
Variability and flares, similar to those observed in late-type stars, have
also been detected in several brown dwarfs. Our results are therefore
consistent with previous observations. 

Two of the very-low mass stars in our sample had been previously detected by
{\it ROSAT}, i.e. S\,Ori~3 and S\,Ori~J053926.8-022614 \citep{mokler02},
with luminosities consistent with the ones derived here. We note however
that S\,Ori~J053926.8-022614 is very close to the $\sim\,$M2 cluster member
SE~94, and its identification is ambiguous: from the XMM-{\it Newton} data
it is not possible to determine which of the two stars is really responsible
for the X-ray emission (see Sect.~\ref{srcid}). At the time of the {\it
ROSAT} observation SE~94 was not known, and the observed X-ray emission was
attributed entirely to S\,Ori~J053926.8-022614. 

To conclude, the XMM-{\it Newton} observations of the cluster around
$\sigma$~Ori reported in this paper are consistent with what is
typically found in SFRs and very young clusters. The coronal temperatures,
metallicities, $L_\mathrm{X}/L_\mathrm{bol}$ ratios and variability of the
PMS stars in the $\sigma$~Ori cluster are similar to those typically
observed in SFRs and clusters in the age range $1-4$~Myr. The median X-ray
luminosity and the detection rate of late-type stars in $\sigma$~Ori,
however, are lower than in clusters of similar age, but this could be due
to significant contamination by non-members. Although we found evidence of
strong absorption in the spectra of some CTTS, and their X-ray luminosities
are typically lower than those of WTTS, we do not find significant
differences in the coronal temperatures and metallicities of CTTS with
respect to WTTS. Our limited data sample does not support the view that CTTS
as a class have all low coronal temperatures and high densities as found
from high-resolution XMM-{\it Newton} and {\it Chandra} observations of the
CTT star TW~Hya. While the latter behaviour has been interpreted as due to
an accretion shock, our sample shows evidence for surface magnetic activity
in at least some CTTS, similarly to what is typically observed in WTT stars.

\begin{acknowledgements}
We thank the anonymous referee for his/her comments, and E. Flaccomio for
useful comments and discussions. EF and RP acknowledge partial support from
Ministero dell'Istruzione, Universit\`a e Ricerca (MIUR). JSF acknowledges
support by the ESA Research Fellows program. This publication makes use of
data products from the Two Micron All Sky Survey, which is a joint project
of the University of Massachusetts and the Infrared Processing and Analysis
Center/California Institute of Technology, and of the Guide Star
Catalogue-II, which is a joint project of the Space Telescope Science
Institute and INAF-Osservatorio Astronomico di Torino. 
\end{acknowledgements}

\bibliographystyle{aa}
\bibliography{biblio}

\begin{thebibliography}{72}
\expandafter\ifx\csname natexlab\endcsname\relax\def\natexlab#1{#1}\fi

\bibitem[{Argiroffi {et~al.}(2004)Argiroffi, Drake, Maggio, Peres, Sciortino,
  \& Harnden}]{argi04}
Argiroffi, C., Drake, J.~J., Maggio, A., {et~al.} 2004, ApJ, 609, 925

\bibitem[{Argiroffi {et~al.}(2005)Argiroffi, Maggio, Peres, Stelzer, \&
  Neuh{\"a}user}]{argi05}
Argiroffi, C., Maggio, A., Peres, G., Stelzer, B., \& Neuh{\"a}user, R. 2005,
  A\&A, 439, 1149

\bibitem[{Baraffe {et~al.}(1998)Baraffe, Chabrier, Allard, \&
  Hauschildt}]{baraffe98}
Baraffe, I., Chabrier, G., Allard, F., \& Hauschildt, P.~H. 1998, A\&A, 337,
  403

\bibitem[{Barrado~y Navascu{\'e}s {et~al.}(2003)Barrado~y Navascu{\'e}s,
  B{\'e}jar, Mundt, Mart{\'\i}n, Rebolo, Zapatero~Osorio, \&
  Bailer-Jones}]{barrado03}
Barrado~y Navascu{\'e}s, D., B{\'e}jar, V.~J.~S., Mundt, R., {et~al.} 2003,
  A\&A, 404, 171

\bibitem[{Barrado~y Navascu{\'e}s \& Mart{\'\i}n(2003)}]{barrado03ha}
Barrado~y Navascu{\'e}s, D. \& Mart{\'\i}n, E.~L. 2003, AJ, 126, 2997

\bibitem[{Barrado~y Navascu{\'e}s {et~al.}(2001)Barrado~y Navascu{\'e}s,
  Zapatero~Osorio, B{\'e}jar, Rebolo, Mart{\'\i}n, Mundt, \&
  Bailer-Jones}]{barrado01}
Barrado~y Navascu{\'e}s, D., Zapatero~Osorio, M.~R., B{\'e}jar, V.~J.~S.,
  {et~al.} 2001, A\&A, 377, L9

\bibitem[{B{\'e}jar {et~al.}(2001)B{\'e}jar, Mart{\'\i}n, Zapatero~Osorio,
  Rebolo, Barrado~y Navascu{\'e}s, {et~al.}}]{bejar01}
B{\'e}jar, V.~J.~S., Mart{\'\i}n, E.~L., Zapatero~Osorio, M.~R., {et~al.} 2001,
  ApJ, 556, 830

\bibitem[{B{\'e}jar {et~al.}(1999)B{\'e}jar, Zapatero~Osorio, \&
  Rebolo}]{bejar99}
B{\'e}jar, V.~J.~S., Zapatero~Osorio, M.~R., \& Rebolo, R. 1999, ApJ, 521, 671

\bibitem[{B{\'e}jar {et~al.}(2004)B{\'e}jar, Zapatero~Osorio, \&
  Rebolo}]{bejar04}
B{\'e}jar, V.~J.~S., Zapatero~Osorio, M.~R., \& Rebolo, R. 2004, AN, 325, 705

\bibitem[{Bergh{\"o}fer {et~al.}(1997)Bergh{\"o}fer, Schmitt, Danner, \&
  Cassinelli}]{bergh97}
Bergh{\"o}fer, T.~W., Schmitt, J.~H.~M.~M., Danner, R., \& Cassinelli, J.~P.
  1997, A\&A, 322, 167

\bibitem[{Brown {et~al.}(1994)Brown, de~Geus, \& de~Zeeuw}]{brown94}
Brown, A.~G.~A., de~Geus, E.~J., \& de~Zeeuw, P.~T. 1994, A\&A, 289, 101

\bibitem[{Burningham {et~al.}(2005)Burningham, Naylor, Littlefair, \&
  Jeffries}]{burn05}
Burningham, B., Naylor, T., Littlefair, S.~P., \& Jeffries, R.~D. 2005, MNRAS,
  356, 1583

\bibitem[{Caballero {et~al.}(2004)Caballero, B{\'e}jar, Rebolo, \&
  Zapatero~Osorio}]{caballero04}
Caballero, J.~A., B{\'e}jar, V.~J.~S., Rebolo, R., \& Zapatero~Osorio, M.~R.
  2004, A\&A, 424, 857

\bibitem[{Cutri {et~al.}(2003)Cutri, Skrutskie, van Dyk, Beichman, Carpenter,
  Chester, Cambresy, Evans, {et~al.}}]{2mass}
Cutri, R.~M., Skrutskie, M.~F., van Dyk, S., {et~al.} 2003, 2MASS All-Sky
  Catalog of Point Sources

\bibitem[{Damiani {et~al.}(1997)Damiani, Maggio, Micela, \&
  Sciortino}]{damiani97}
Damiani, F., Maggio, A., Micela, G., \& Sciortino, S. 1997, ApJ, 483, 350

\bibitem[{ESA(1997)}]{esa97}
ESA. 1997, The Hipparcos and Tycho catalogues, ESA SP-1200

\bibitem[{Favata {et~al.}(2003)Favata, Giardino, Micela, Sciortino, \&
  Damiani}]{favata03sfr}
Favata, F., Giardino, G., Micela, G., Sciortino, S., \& Damiani, F. 2003, A\&A,
  403, 187

\bibitem[{Feigelson {et~al.}(2002)Feigelson, Broos, Gaffney, Garmire,
  Hillenbrand, Pravdo, Townsley, \& Tsuboi}]{feigel02}
Feigelson, E.~D., Broos, P., Gaffney, J.~A., I., {et~al.} 2002, ApJ, 574, 258

\bibitem[{Feigelson {et~al.}(2003)Feigelson, Gaffney, Garmire, Hillenbrand, \&
  Townsley}]{feigel03}
Feigelson, E.~D., Gaffney, J.~A., I., Garmire, G., Hillenbrand, L.~A., \&
  Townsley, L. 2003, ApJ, 584, 911

\bibitem[{Feigelson \& Lawson(2004)}]{feigel04}
Feigelson, E.~D. \& Lawson, W.~A. 2004, ApJ, 614, 267

\bibitem[{Feigelson \& Montmerle(1999)}]{feigel99}
Feigelson, E.~D. \& Montmerle, T. 1999, ARA\&A, 37, 363

\bibitem[{Flaccomio {et~al.}(2003{\natexlab{a}})Flaccomio, Damiani, Micela,
  Sciortino, Harnden, Murray, \& Wolk}]{flaccomio03o1}
Flaccomio, E., Damiani, F., Micela, G., {et~al.} 2003{\natexlab{a}}, ApJ, 582,
  382

\bibitem[{Flaccomio {et~al.}(2003{\natexlab{b}})Flaccomio, Damiani, Micela,
  Sciortino, Harnden, Murray, \& Wolk}]{flaccomio03o2}
Flaccomio, E., Damiani, F., Micela, G., {et~al.} 2003{\natexlab{b}}, ApJ, 582,
  398

\bibitem[{Flaccomio {et~al.}(2003{\natexlab{c}})Flaccomio, Micela, \&
  Sciortino}]{flaccomio03}
Flaccomio, E., Micela, G., \& Sciortino, S. 2003{\natexlab{c}}, A\&A, 397, 611

\bibitem[{Flaccomio {et~al.}(2003{\natexlab{d}})Flaccomio, Micela, \&
  Sciortino}]{flaccomio03ev}
Flaccomio, E., Micela, G., \& Sciortino, S. 2003{\natexlab{d}}, A\&A, 402, 277

\bibitem[{Gagn{\'e} {et~al.}(2004)Gagn{\'e}, Skinner, \& Daniel}]{gagne04}
Gagn{\'e}, M., Skinner, S.~L., \& Daniel, K.~J. 2004, ApJ, 613, 393

\bibitem[{Garmire {et~al.}(2000)Garmire, Feigelson, Broos, Hillenbrand, Pravdo,
  Townsley, \& Tsuboi}]{garmire00}
Garmire, G., Feigelson, E.~D., Broos, P., {et~al.} 2000, AJ, 120, 1426

\bibitem[{Getman {et~al.}(2002)Getman, Feigelson, Townsley, Bally, Lada, \&
  Reipurth}]{getman02}
Getman, K.~V., Feigelson, E.~D., Townsley, L., {et~al.} 2002, ApJ, 575, 354

\bibitem[{Hasinger {et~al.}(2001)Hasinger, Altieri, Arnaud, Barcons, Bergeron,
  {et~al.}}]{hasinger01}
Hasinger, G., Altieri, B., Arnaud, M., {et~al.} 2001, A\&A, 365, L45

\bibitem[{Imanishi {et~al.}(2001)Imanishi, Tsujimoto, \& Koyama}]{imanishi01bd}
Imanishi, K., Tsujimoto, M., \& Koyama, K. 2001, ApJ, 563, 361

\bibitem[{Imanishi {et~al.}(2002)Imanishi, Tsujimoto, \& Koyama}]{imanishi02}
Imanishi, K., Tsujimoto, M., \& Koyama, K. 2002, ApJ, 572, 300

\bibitem[{Kastner {et~al.}(2004)Kastner, Huenemorder, Schulz, Canizares, Li, \&
  Weintraub}]{kastner04}
Kastner, J.~H., Huenemorder, D.~P., Schulz, N.~S., {et~al.} 2004, ApJ, 605, L49

\bibitem[{Kastner {et~al.}(2002)Kastner, Huenemorder, Schulz, Canizares, \&
  Weintraub}]{kastner02}
Kastner, J.~H., Huenemorder, D.~P., Schulz, N.~S., Canizares, C.~R., \&
  Weintraub, D.~A. 2002, ApJ, 567, 434

\bibitem[{Kenyon {et~al.}(2005)Kenyon, Jeffries, Naylor, Oliveira, \&
  Maxted}]{kenyon05}
Kenyon, M.~J., Jeffries, R.~D., Naylor, T., Oliveira, J.~M., \& Maxted,
  P.~F.~L. 2005, MNRAS, 356, 89

\bibitem[{Kenyon \& Hartmann(1995)}]{kh95}
Kenyon, S.~J. \& Hartmann, L. 1995, ApJS, 101, 117

\bibitem[{Lavalley {et~al.}(1992)Lavalley, Isobe, \& Feigelson}]{asurv}
Lavalley, M., Isobe, T., \& Feigelson, E. 1992, in Astronomical Data Analysis
  Software and Systems, ed. D.~M. Worrall {et~al.}, ASP Conf. Ser. No.~25, 245

\bibitem[{Lee(1968)}]{lee68}
Lee, T.~A. 1968, ApJ, 152, 913

\bibitem[{Leggett {et~al.}(2001)Leggett, Allard, Geballe, Hauschildt, \&
  Schweitzer}]{leggett01}
Leggett, S.~K., Allard, F., Geballe, T.~R., Hauschildt, P.~H., \& Schweitzer,
  A. 2001, ApJ, 548, 908

\bibitem[{Mart{\'\i}n {et~al.}(2001)Mart{\'\i}n, Zapatero~Osorio, Barrado~y
  Navascu{\'e}s, B{\'e}jar, \& Rebolo}]{martin01}
Mart{\'\i}n, E.~L., Zapatero~Osorio, M.~R., Barrado~y Navascu{\'e}s, D.,
  B{\'e}jar, V.~J.~S., \& Rebolo, R. 2001, ApJ, 558, L117

\bibitem[{Mokler \& Stelzer(2002)}]{mokler02}
Mokler, F. \& Stelzer, B. 2002, A\&A, 391, 1025

\bibitem[{Muzerolle {et~al.}(2003)Muzerolle, Hillenbrand, Calvet, Brice{\~n}o,
  \& Hartmann}]{muzerolle03}
Muzerolle, J., Hillenbrand, L., Calvet, N., Brice{\~n}o, C., \& Hartmann, L.
  2003, ApJ, 266, 592

\bibitem[{Neuh{\"a}user {et~al.}(1999)Neuh{\"a}user, Brice{\~n}o, Comer{\'o}n,
  Hearty, Mart{\'\i}n, {et~al.}}]{neuh99}
Neuh{\"a}user, R., Brice{\~n}o, C., Comer{\'o}n, F., {et~al.} 1999, A\&A, 343,
  883

\bibitem[{Oliveira {et~al.}(2002)Oliveira, Jeffries, Kenyon, Thompson, \&
  Naylor}]{oliveira02}
Oliveira, J.~M., Jeffries, R.~D., Kenyon, M.~J., Thompson, S.~A., \& Naylor, T.
  2002, A\&A, 382, L22

\bibitem[{Oliveira {et~al.}(2004)Oliveira, Jeffries, \& van Loon}]{oliveira04}
Oliveira, J.~M., Jeffries, R.~D., \& van Loon, J.~T. 2004, MNRAS, 347, 1327

\bibitem[{Oliveira \& van Loon(2004)}]{olivloon04}
Oliveira, J.~M. \& van Loon, J.~T. 2004, A\&A, 418, 663

\bibitem[{Ozawa {et~al.}(2005)Ozawa, Grosso, \& Montmerle}]{ozawa05}
Ozawa, H., Grosso, N., \& Montmerle, T. 2005, A\&A, 429, 963

\bibitem[{Pallavicini {et~al.}(2005)Pallavicini, Franciosini, Maggio, Scelsi,
  \& Sanz-Forcada}]{pallavic05asr}
Pallavicini, R., Franciosini, E., Maggio, A., Scelsi, L., \& Sanz-Forcada, J.
  2005, Adv. Sp. Res., in press

\bibitem[{Pallavicini {et~al.}(1981)Pallavicini, Golub, Rosner, Vaiana, Ayres,
  \& Linsky}]{pallavic81}
Pallavicini, R., Golub, L., Rosner, R., {et~al.} 1981, ApJ, 248, 279

\bibitem[{Preibisch \& Zinnecker(2001)}]{preibisch01}
Preibisch, T. \& Zinnecker, H. 2001, AJ, 122, 866

\bibitem[{Preibisch \& Zinnecker(2002)}]{preibisch02}
Preibisch, T. \& Zinnecker, H. 2002, AJ, 123, 1613

\bibitem[{Preibisch \& Zinnecker(2004)}]{preibisch04}
Preibisch, T. \& Zinnecker, H. 2004, A\&A, 422, 1001

\bibitem[{Ram{\'\i}rez {et~al.}(2004)Ram{\'\i}rez, Rebull, Stauffer, Hearty,
  Hillenbrand, {et~al.}}]{ramirez04}
Ram{\'\i}rez, S.~V., Rebull, L., Stauffer, J., {et~al.} 2004, AJ, 127, 2659

\bibitem[{Randich \& Schmitt(1995)}]{randsch95}
Randich, S. \& Schmitt, J.~H.~M.~M. 1995, A\&A, 298, 115

\bibitem[{Sanz-Forcada {et~al.}(2004)Sanz-Forcada, Franciosini, \&
  Pallavicini}]{sanz04sig}
Sanz-Forcada, J., Franciosini, E., \& Pallavicini, R. 2004, A\&A, 421, 715

\bibitem[{Scelsi {et~al.}(2005)Scelsi, Maggio, Peres, \&
  Pallavicini}]{scelsi05}
Scelsi, L., Maggio, A., Peres, G., \& Pallavicini, R. 2005, A\&A, 432, 671

\bibitem[{Schmitt {et~al.}(2005)Schmitt, Robrade, Ness, Favata, \&
  Stelzer}]{schmitt05}
Schmitt, J.~H.~M.~M., Robrade, J., Ness, J.-U., Favata, F., \& Stelzer, B.
  2005, A\&A, 432, L35

\bibitem[{Scholz \& Eisl{\"o}ffel(2004)}]{se04}
Scholz, A. \& Eisl{\"o}ffel, J. 2004, A\&A, 419, 249

\bibitem[{Sherry {et~al.}(2004)Sherry, Walter, \& Wolk}]{sww04}
Sherry, W.~H., Walter, F.~M., \& Wolk, S.~J. 2004, AJ, 128, 2316

\bibitem[{Siess {et~al.}(2000)Siess, Dufour, \& Forestini}]{siess00}
Siess, L., Dufour, E., \& Forestini, M. 2000, A\&A, 358, 593

\bibitem[{Skinner {et~al.}(2003)Skinner, Gagn{\'e}, \& Belzer}]{skinner03}
Skinner, S., Gagn{\'e}, M., \& Belzer, E. 2003, ApJ, 598, 375

\bibitem[{Stassun {et~al.}(2004)Stassun, Ardila, Barsony, Basri, \&
  Mathieu}]{stassun04}
Stassun, K.~G., Ardila, D.~R., Barsony, M., Basri, G., \& Mathieu, R.~D. 2004,
  AJ, 127, 3537

\bibitem[{Stelzer {et~al.}(2004)Stelzer, Micela, \&
  Neuh{\"a}user}]{stelzer04cha}
Stelzer, B., Micela, G., \& Neuh{\"a}user, R. 2004, A\&A, 423, 1029

\bibitem[{Stelzer \& Neuh{\"a}user(2001)}]{stelzer01}
Stelzer, B. \& Neuh{\"a}user, R. 2001, A\&A, 377, 538

\bibitem[{Stelzer {et~al.}(1999)Stelzer, Neuh{\"a}user, Casanova, \&
  Montmerle}]{stelzer99}
Stelzer, B., Neuh{\"a}user, R., Casanova, S., \& Montmerle, T. 1999, A\&A, 344,
  154

\bibitem[{Stelzer \& Schmitt(2004)}]{stelzer04}
Stelzer, B. \& Schmitt, J.~H.~M.~M. 2004, A\&A, 418, 687

\bibitem[{Tozzi {et~al.}(2001)Tozzi, Rosati, Nonino, Bergeron, Borgani,
  {et~al.}}]{tozzi01}
Tozzi, P., Rosati, P., Nonino, M., {et~al.} 2001, ApJ, 562, 42

\bibitem[{Walter {et~al.}(1997)Walter, Wolk, Freyberg, \& Schmitt}]{walter97}
Walter, F.~M., Wolk, S.~J., Freyberg, M., \& Schmitt, J.~H.~M.~M. 1997, Mem.
  Soc. Astr. It., 68, 1081

\bibitem[{Warren \& Hesser(1977)}]{wh77}
Warren, W.~H., J. \& Hesser, J.~E. 1977, ApJS, 34, 115

\bibitem[{Weaver \& Babcock(2004)}]{weaver04}
Weaver, W.~B. \& Babcock, A. 2004, PASP, 116, 1035

\bibitem[{Wolk(1996)}]{wolk96}
Wolk, S.~J. 1996, PhD thesis, Univ. New York at Stony Brook

\bibitem[{Zapatero~Osorio {et~al.}(2000)Zapatero~Osorio, B{\'e}jar,
  Mart{\'\i}n, Rebolo, Barrado~y Navascu{\'e}s, Bayler-Jones, \&
  Mundt}]{zapat00}
Zapatero~Osorio, M.~R., B{\'e}jar, V.~J.~S., Mart{\'\i}n, E.~L., {et~al.} 2000,
  Science, 290, 103

\bibitem[{Zapatero~Osorio {et~al.}(2002)Zapatero~Osorio, B{\'e}jar, Pavlenko,
  Rebolo, Allende~Prieto, Mart{\'\i}n, \& Garc{\'\i}a~L{\'o}pez}]{zapat02}
Zapatero~Osorio, M.~R., B{\'e}jar, V.~J.~S., Pavlenko, Y., {et~al.} 2002, A\&A,
  384, 937

\end{thebibliography}

\appendix
\section{X-ray and optical properties of cluster members and candidates} 
\small
\begin{longtable}{rcrrlcrcclll}
\caption{\label{detmem} X-ray sources with at least one cluster member or
candidate within 5$\arcsec$. NX is a running identification number for the
X-ray sources. The column labeled ``Sign.'' indicates the significance of
detection. Optical identifications are from the following sources: 4771-...,
r05..., p05... $=$ \citet{wolk96}; SE $=$ \citet{se04}; SWW $=$\citet{sww04}; 
S\,Ori, J05... $=$ \citet{bejar99,bejar01,bejar04,zapat02,caballero04}: note
that we have dropped the S\,Ori prefix in front of the J05... names; K $=$
\citet{kenyon05}. The flag in the column labeled "C/W" indicates whether the
star is a probable/possible CTTS or WTTS} \\
\hline\hline
\noalign{\smallskip}
NX& RA$_\mathrm{X}$ \hspace{24pt} DEC$_\mathrm{X}$& Sign.& Count
rate& Optical ID& 
  $\Delta r$& $I$\ \ \ \ & $R\!-\!I$& $I\!-\!J$& SpT$^{\,a}$& $\log
L_\mathrm{X}^{\,b}$ \ \ & C/\\
 & (J2000)& & (cts/ks)& & ($\,\arcsec\,$)& & & & & & W\\
\noalign{\smallskip} \hline \noalign{\smallskip}
\endfirsthead
\caption{continued.} \\
\hline\hline
\noalign{\smallskip}
NX& RA$_\mathrm{X}$ \hspace{25pt} DEC$_\mathrm{X}$& Sign.& Count
rate& Optical ID&
  $\Delta r$& $I$\ \ \ \ & $R\!-\!I$& $I\!-\!J$& SpT$^{\,a}$& $\log
L_\mathrm{X}^{\,b}$ \ \ & C/\\
& (J2000)& & (cts/ks)& & ($\,\arcsec\,$)& & & & & & W\\
\noalign{\smallskip} \hline \noalign{\smallskip}
\endhead
\noalign{\smallskip} \hline 
\endfoot
\noalign{\smallskip} \hline \noalign{\smallskip}
\multicolumn{12}{l}{$a$: spectroscopically-determined spectral types from
the literature are marked with an asterisk. The other spectral types have
been}\\
\multicolumn{12}{l}{\phantom{b:} estimated from the $R-I$, $V-I$ or $I-J$
colours using the relations by \citet{kh95} and \citet{leggett01} }\\
\multicolumn{12}{l}{$b$: unabsorbed X-ray luminosity in the $0.3-8$ keV
band. Values marked with a $\dag$\ have been derived from the PN or MOS1
spectral fits }\\
\multicolumn{12}{l}{\phantom{b:} reported in Table~\ref{tabfits} and in
Paper I}\\
\multicolumn{12}{l}{$c$: source \#3 in Paper I}\\
\multicolumn{12}{l}{$d$: source \#4 in Paper I}\\
\multicolumn{12}{l}{$e$: source detected on the PN only. The reported count
rates are PN count rates. $L_\mathrm{X}$ was derived using the conversion
factor for PN }\\
\multicolumn{12}{l}{\phantom{b:} (see Sect.~\ref{Lx})}\\
\endlastfoot
  1& 5:37:51.62 \ \ $-$2:35:23.6&  21.5& $ 10.55 \pm 0.94$& SWW 125         & 2.48& 13.34& 1.22& 1.45& M2      & 30.02       &   \\
  2& 5:37:53.03 \ \ $-$2:33:33.7& 104.9& $148.74 \pm 3.45$& 4771-0775       & 1.65& 10.69& 0.42& 0.70& K0$^*$  & 31.13$^\dag$& C?\\
  3& 5:37:54.44 \ \ $-$2:39:27.8& 122.0& $189.27 \pm 3.81$& 4771-0921       & 2.22& 10.04& 0.51& 0.78& K0$^*$  & 31.24$^\dag$& W?\\
  4& 5:38:00.85 \ \ $-$2:45:09.4&  17.2& $  9.68 \pm 0.97$& SWW 140         & 2.93& 14.54& 1.77& 1.81& M4      & 29.98       &   \\
  6& 5:38:06.52 \ \ $-$2:28:48.9&   5.4& $  0.67 \pm 0.18$& 4771-0950       & 1.11& 10.70&     & 0.61& F7      & 28.82       & W?\\
  7& 5:38:06.77 \ \ $-$2:30:22.9&  50.8& $ 19.05 \pm 0.86$& SWW 113         & 1.02& 13.14& 0.94& 1.38& M0      & 30.77$^\dag$& C \\
  8& 5:38:07.94 \ \ $-$2:31:31.0& 102.3& $ 49.95 \pm 1.22$& 4771-0854       & 0.38& 11.45& 0.64& 0.88& K4      & 30.66$^\dag$&   \\
  9& 5:38:08.33 \ \ $-$2:35:57.5&  20.4& $  3.42 \pm 0.32$& SWW 41          & 1.29& 13.76& 1.32& 1.62& M2      & 29.53       & W?\\
 11& 5:38:13.22 \ \ $-$2:26:07.9&   9.1& $  1.63 \pm 0.29$& SE 3            & 1.39& 14.10& 1.57& 1.62& M3      & 29.20       &   \\
 17& 5:38:17.88 \ \ $-$2:40:49.4&   9.1& $  0.97 \pm 0.17$& J053817.8-024050& 0.63& 14.98& 1.83& 1.78& M4      & 28.98       &   \\
 19& 5:38:18.39 \ \ $-$2:35:38.0&   9.9& $  0.95 \pm 0.16$& J053818.2-023539& 0.97& 17.96& 2.23& 2.51& M5      & 28.97       &   \\
 20& 5:38:20.29 \ \ $-$2:38:01.3&  15.7& $  1.37 \pm 0.18$& J053820.1-023802& 0.42& 14.33& 1.69& 1.75& M4.0$^*$& 29.13       & W \\
 25& 5:38:23.35 \ \ $-$2:44:14.6&   8.6& $  0.85 \pm 0.17$& J053823.3-024414& 1.10& 15.19& 1.73& 1.73& M3      & 28.92       &   \\
 26& 5:38:23.59 \ \ $-$2:41:33.9&   9.1& $  0.96 \pm 0.19$& J053823.6-024132& 2.28& 14.93& 1.81& 1.64& M4      & 28.98       &   \\
 31& 5:38:26.52 \ \ $-$2:34:28.7&  31.7& $  4.68 \pm 0.29$& 4771-1021       & 0.46& 12.10& 0.50& 0.68& K2      & 29.66       &   \\
 32& 5:38:27.22 \ \ $-$2:45:09.1&  24.2& $  4.76 \pm 0.38$& 4771-0041       & 2.03& 12.82& 0.82& 0.87& K7.0$^*$& 29.67       & C \\
 33& 5:38:27.62 \ \ $-$2:35:02.6&   7.0& $  1.54 \pm 0.40$& J053827.5-023504& 1.60& 14.41& 1.42& 1.58& M3.5$^*$& 29.18       & C \\
 34& 5:38:27.75 \ \ $-$2:43:00.5&  39.7& $  8.69 \pm 0.49$& SWW 87          & 1.36& 13.67& 1.33& 1.48& M3      & 29.96$^\dag$&   \\
 39& 5:38:29.19 \ \ $-$2:36:02.7&  62.6& $ 12.51 \pm 0.45$& SWW 177         & 0.35& 14.04& 1.18& 1.40& M2      & 30.08$^\dag$&   \\
 44& 5:38:31.66 \ \ $-$2:35:14.8&  28.3& $  3.89 \pm 0.29$& r053831-0235    & 0.18& 13.49& 1.11& 1.97& M0.0$^*$& 29.58       & W \\
 49& 5:38:32.97 \ \ $-$2:35:39.6&  74.7& $ 18.73 \pm 0.54$& r053832-0235b   & 0.59& 12.88& 0.83& 1.34& K7      & 30.27$^\dag$&   \\
 52& 5:38:33.36 \ \ $-$2:36:17.6&  22.1& $  3.17 \pm 0.25$& SWW 130         & 1.42& 13.47& 1.26& 1.42& M2      & 29.49       &   \\
 53& 5:38:33.90 \ \ $-$2:44:13.8&  30.9& $  7.72 \pm 0.55$& 4771-1095       & 1.84& 11.47& 0.81& 1.34& K5$^*$  & 29.88       & C \\
 54& 5:38:34.03 \ \ $-$2:36:37.5&  10.5& $  1.12 \pm 0.19$& r053833-0236    & 1.84& 13.72& 1.52& 1.74& M3.5$^*$& 29.04       & C \\
 55& 5:38:34.20 \ \ $-$2:34:15.5&  39.7& $  5.86 \pm 0.35$& HD 294272       & 4.54&  8.57&     &     & B9.5$^*$& 29.76       &   \\
 60& 5:38:35.74 \ \ $-$2:31:52.2&  59.6& $ 16.05 \pm 0.65$& r053835-0231    & 2.83& 12.45& 0.79& 1.15& K5$^*$  & 30.20       & W \\
 61& 5:38:35.88 \ \ $-$2:43:49.5&  40.3& $  9.16 \pm 0.48$& 4771-1026       & 2.14& 11.99& 0.72& 1.55& K3$^*$  & 30.38$^\dag$& C \\
 62& 5:38:36.01 \ \ $-$2:30:43.3& 209.2& $143.97 \pm 1.82$& 4771-1097       & 0.67& 12.47& 0.81& 1.23& K6.0$^*$& 31.08$^\dag$& W \\
 64& 5:38:38.28 \ \ $-$2:36:38.2& 123.3& $ 37.57 \pm 0.74$& r053838-0236    & 0.60& 12.38& 0.86& 1.22& K8.0$^*$& 30.62$^\dag$& W \\
 65& 5:38:38.54 \ \ $-$2:34:55.5& 252.1& $138.80 \pm 1.43$& 4771-1147$^{\,c}$& 0.75& 10.88& 0.65& 0.97& K0$^*$ & 31.07$^\dag$& W \\
 67& 5:38:39.13 \ \ $-$2:28:00.4&   9.7& $  1.29 \pm 0.21$& SE 70           & 2.60& 16.99& 1.78& 1.72& M4      & 29.11       &   \\
   &                            &      &                  & S\,Ori 68       & 4.72& 23.78&     & 3.60& L5.0$^*$&             & W \\
 68& 5:38:39.76 \ \ $-$2:40:19.5&  10.4& $  0.98 \pm 0.15$& p053839-0240    & 0.95& 15.50& 1.77& 1.75& M4      & 28.98       &   \\
 69& 5:38:40.35 \ \ $-$2:30:17.5&  62.6& $ 16.11 \pm 0.57$& r053840-0230    & 1.10& 12.80& 0.94& 1.29& M0.0$^*$& 30.52$^\dag$& C \\
 70& 5:38:41.34 \ \ $-$2:37:22.3& 224.3& $107.17 \pm 1.24$& r053841-0237$^{\,d}$& 0.78& 12.77& 0.87& 1.31& K3$^*$& 30.99$^\dag$& C\\
 71& 5:38:41.46 \ \ $-$2:36:43.7&   9.2& $  1.08 \pm 0.17$& p053841-0236    & 0.81& 14.52& 1.36& 1.53& M3      & 29.03       &   \\
 75& 5:38:43.38 \ \ $-$2:32:00.5&  12.5& $  0.67 \pm 0.11$& SWW 144         & 0.88& 14.12& 1.75& 1.88& M4      & 28.82       &   \\
 76& 5:38:43.63 \ \ $-$2:33:25.7&  83.1& $ 19.35 \pm 0.55$& SWW 36          & 0.38& 13.09& 1.06& 1.37& M1      & 30.28$^\dag$&   \\
 78& 5:38:44.27 \ \ $-$2:40:19.7& 137.5& $ 48.80 \pm 0.91$& 4771-1051       & 0.77& 12.33& 0.79& 0.97& K7.5$^*$& 30.68$^\dag$& W \\
 79& 5:38:44.36 \ \ $-$2:32:33.7& 162.5& $ 64.62 \pm 1.02$& 4771-1055       & 0.34& 12.04& 0.86& 1.16& K8      & 30.81$^\dag$&   \\
 80& 5:38:44.86 \ \ $-$2:36:00.2& 479.2& $439.86 \pm 2.47$& $\sigma$ Ori AB & 0.24&  4.07&     &     & O9.5$^*$& 31.74$^\dag$&   \\
 81& 5:38:45.40 \ \ $-$2:41:59.7&  72.1& $ 19.21 \pm 0.62$& SWW 97          & 1.00& 13.46& 1.16& 1.47& M1      & 30.38$^\dag$&   \\
 82& 5:38:46.10 \ \ $-$2:45:25.1&   5.4& $  0.73 \pm 0.18$& J053845.9-024523& 1.98& 15.52& 1.84& 1.96& M4      & 28.86       &   \\
 84& 5:38:47.26 \ \ $-$2:35:39.9& 312.7& $199.42 \pm 1.69$& $\sigma$ Ori E  & 0.81&  6.49&     &     & B2p$^*$ & 31.46$^\dag$&   \\
 87& 5:38:47.56 \ \ $-$2:35:24.7&  25.4& $  5.89 \pm 0.58$& SWW 78          & 0.54& 12.86& 0.99& 1.12& M1      & 29.76       &   \\
 88& 5:38:47.99 \ \ $-$2:37:19.7&  25.0& $  3.99 \pm 0.26$& SWW 102         & 0.55& 13.02& 1.18& 1.00& K5$^*$  & 29.59       & W \\
 89& 5:38:48.05 \ \ $-$2:30:39.3& 5.5& $2.15 \pm 0.54$& S\,Ori 6$^{\,e}$    & 4.88& 15.53& 2.00& 1.88& M5      & 28.83       &   \\
 90& 5:38:48.18 \ \ $-$2:27:13.3& 166.7& $106.85 \pm 1.72$& 4771-0899       & 1.14& 11.35& 0.82& 1.19& K7.0$^*$& 31.21$^\dag$& W \\
 92& 5:38:48.76 \ \ $-$2:36:16.7&  59.5& $ 13.60 \pm 0.48$& SWW 35          & 0.51& 13.37& 0.87& 1.26& M1      & 30.20$^\dag$&   \\
 93& 5:38:49.26 \ \ $-$2:38:23.5&  93.0& $ 22.49 \pm 0.58$& r053849-0238    & 1.28& 12.96& 1.03& 1.57& M0.5$^*$& 30.31$^\dag$& C \\
 94& 5:38:49.27 \ \ $-$2:41:25.3&  13.9& $  1.00 \pm 0.16$& SWW 205         & 0.78& 13.23& 1.50& 1.56& M3      & 28.99       &   \\
 97& 5:38:49.90 \ \ $-$2:41:23.4&  15.2& $  1.08 \pm 0.15$& SWW 200         & 0.75& 14.21& 1.37& 1.46& M3      & 29.03       &   \\
 98& 5:38:50.19 \ \ $-$2:37:34.4&  28.0& $  3.68 \pm 0.25$& SWW 18          & 1.38& 14.83& 1.69& 1.78& M4      & 29.56       &   \\
100& 5:38:50.54 \ \ $-$2:26:47.0&   7.6& $  1.04 \pm 0.23$& SE 34           & 1.08& 14.10& 1.48& 1.60& M3      & 29.01       &   \\
102& 5:38:51.53 \ \ $-$2:36:20.7&  39.7& $  5.03 \pm 0.29$& r053851-0236    & 0.25& 13.38& 0.87& 0.94& K9      & 29.69       & C \\
104& 5:38:52.07 \ \ $-$2:46:43.9&  52.2& $ 18.84 \pm 0.79$& 4771-0080       & 0.57& 12.78& 0.80& 1.26& K7      & 30.19$^\dag$& W \\
105& 5:38:53.15 \ \ $-$2:38:53.1&  17.6& $  1.72 \pm 0.18$& SWW 166         & 0.57& 12.78& 1.07& 1.16& M1      & 29.23       &   \\
106& 5:38:53.20 \ \ $-$2:43:53.5&  55.5& $ 15.01 \pm 0.60$& SWW 47          & 1.24& 13.49& 1.08& 1.26& M1      & 30.12$^\dag$& W \\
109& 5:38:53.63 \ \ $-$2:33:23.0& 159.2& $ 69.21 \pm 1.20$& 4771-1049       & 2.39& 11.57& 0.73& 0.96& K5$^*$  & 30.86$^\dag$& W \\
110& 5:38:54.19 \ \ $-$2:49:29.4&  50.8& $ 53.52 \pm 2.31$& 4771-0119       & 0.41& 11.73& 0.58& 0.90& K2$^*$  & 30.67$^\dag$& W \\
111& 5:38:55.13 \ \ $-$2:28:58.9&   5.9& $  0.69 \pm 0.16$& SE 77           & 1.79& 15.50& 1.76& 1.70& M4      & 28.83       &   \\
117& 5:38:59.23 \ \ $-$2:47:14.1&  23.7& $  5.83 \pm 0.48$& r053859-0247    & 0.82& 12.36& 0.69& 1.04& K5      & 29.76       &   \\
118& 5:38:59.32 \ \ $-$2:33:49.3&   6.8& $  0.25 \pm 0.07$& SWW 227         & 2.09& 14.83& 1.26& 1.94& M0      & 28.39       & C?\\
122& 5:39:00.65 \ \ $-$2:39:39.9&  49.5& $  9.95 \pm 0.44$& 4771-1056       & 1.01& 12.43& 0.49& 0.77& K1$^*$  & 29.97$^\dag$& W?\\
124& 5:39:01.24 \ \ $-$2:36:39.2&  11.1& $  0.80 \pm 0.12$& K 9             & 0.38& 15.11& 1.56& 1.59& M3      & 28.90       &   \\
126& 5:39:02.89 \ \ $-$2:29:55.7&  30.2& $  5.32 \pm 0.36$& SWW 28          & 0.53& 14.21& 1.51& 1.60& M3      & 29.72       &   \\
131& 5:39:05.30 \ \ $-$2:33:00.2&   7.2& $  0.34 \pm 0.09$& SWW 175         & 0.64& 15.02& 1.55& 1.63& M3      & 28.52       &   \\
132& 5:39:05.49 \ \ $-$2:32:30.7&  52.2& $ 12.41 \pm 0.52$& 4771-1075       & 0.40& 12.66& 0.87& 1.11& K7.0$^*$& 30.10$^\dag$& W \\
137& 5:39:07.69 \ \ $-$2:28:21.5&  11.5& $  1.76 \pm 0.26$& r053907-0228    & 1.92& 14.37& 1.45& 1.49& M3.0$^*$& 29.24       & W \\
138& 5:39:07.77 \ \ $-$2:32:39.7&  56.0& $ 19.63 \pm 0.83$& 4771-1092       & 1.24& 12.63& 0.81& 1.33& K5$^*$  & 30.50$^\dag$& C \\
140& 5:39:09.21 \ \ $-$2:39:59.4&   5.0& $  0.31 \pm 0.09$& S\,Ori 25       & 2.87& 17.16& 2.17& 2.27& M7.5$^*$& 28.48       & C?\\
144& 5:39:11.66 \ \ $-$2:31:06.0&  26.6& $ 13.68 \pm 0.93$& SWW 195         & 0.98& 13.45& 0.95& 1.46& K5$^*$  & 30.13       & C \\
145& 5:39:11.74 \ \ $-$2:36:03.3&  60.8& $ 15.85 \pm 0.59$& 4771-1038       & 0.54& 12.93& 0.95& 1.31& K8$^*$  & 30.32$^\dag$& W \\
147& 5:39:12.74 \ \ $-$2:30:07.7&  12.0& $  2.09 \pm 0.36$& SWW 203         & 4.95& 14.69& 2.00& 2.08& M5      & 29.31       &   \\
149& 5:39:14.60 \ \ $-$2:28:33.4&  36.5& $ 10.90 \pm 0.62$& J053914.5-022834& 0.55& 14.85& 1.52& 1.51& M3.5$^*$& 30.17$^\dag$& W \\
150& 5:39:15.18 \ \ $-$2:31:37.9&  39.4& $  9.86 \pm 0.53$& HD 37564        & 0.52&  8.17&     & 0.19& A8$^*$  & 29.94$^\dag$&   \\
154& 5:39:17.41 \ \ $-$2:25:44.2&   7.7& $  1.68 \pm 0.33$& SE 51           & 2.28& 14.18& 1.15& 1.28& M2      & 29.22       & C?\\
156& 5:39:18.22 \ \ $-$2:29:28.7&  58.9& $ 23.27 \pm 0.90$& 4771-0598       & 0.85& 11.41& 0.47& 0.69& K1      & 30.28$^\dag$&   \\
157& 5:39:18.97 \ \ $-$2:30:55.6&  15.1& $  2.98 \pm 0.38$& 4771-0910       & 2.53& 12.58& 0.84& 1.18& K3$^*$  & 29.47       & C \\
159& 5:39:20.62 \ \ $-$2:27:37.8&  26.1& $  7.93 \pm 0.60$& J053920.5-022737& 1.58& 13.52& 1.29& 1.37& M2.0$^*$& 29.89       & W \\
160& 5:39:21.08 \ \ $-$2:30:33.8&  24.3& $  5.77 \pm 0.46$& S\,Ori 3        & 0.44& 15.40& 1.98& 2.11& M6      & 29.75       &   \\
161& 5:39:23.02 \ \ $-$2:33:34.3&  23.5& $  4.69 \pm 0.39$& r053923-0233    & 1.56& 14.19& 1.21& 1.36& M2.0$^*$& 29.66       & W \\
164& 5:39:24.48 \ \ $-$2:34:02.7&  16.9& $  3.35 \pm 0.36$& SWW 127         & 1.39& 14.28& 1.25& 1.30& M2      & 29.52       &   \\
165& 5:39:25.33 \ \ $-$2:38:24.3&  11.1& $  1.37 \pm 0.22$& SWW 135         & 2.34& 12.83& 0.95& 1.52& M0      & 29.13       & C?\\
167& 5:39:26.73 \ \ $-$2:26:17.4&   9.6& $  2.94 \pm 0.56$& SE 94           & 3.18& 14.81& 1.17& 2.37& M2      & 29.46        &   \\
   &                            &      &                  & J053926.8-022614& 3.92& 18.66&     & 1.41& M6      &             &   \\
170& 5:39:32.68 \ \ $-$2:39:45.2&  42.8& $ 16.59 \pm 0.85$& r053932-0239    & 1.23& 12.23& 0.81& 1.41& M0      & 30.19$^\dag$&   \\
172& 5:39:36.77 \ \ $-$2:42:17.4&  72.8& $ 63.33 \pm 2.16$& r053936-0242    & 1.99&  9.11& 0.40& 0.65& G5$^*$  & 30.96$^\dag$& W \\
173& 5:39:37.28 \ \ $-$2:26:54.7&  10.7& $  4.40 \pm 0.75$& SWW 163         & 2.62& 12.98& 0.91& 1.28& M0      & 29.64       &   \\
174& 5:39:39.96 \ \ $-$2:33:16.8&  40.3& $ 19.85 \pm 1.04$& V603 Ori        & 1.01& 14.40&     & 2.18& M2      & 31.09$^\dag$& C \\
\end{longtable}
\normalsize

\begin{longtable}{lccrccllrr}
\caption{\label{upplim}$3\sigma$ upper limits for undetected late-type cluster members
or candidates. Optical identifications are from the same sources as in
Table~\ref{detmem}, with the addition of two stars from \citet[B]{burn05}
}\\
\hline\hline
\noalign{\smallskip}
Name& RA& DEC& $I$\ \ \ \ & $R-I$& $I-J$& SpT$^{\,a}$& C/& Count rate$^{\,b}$&
$\log L_\mathrm{X}^{\,c}$\ \ \\
&\multicolumn{2}{c}{(J2000)}& & & & & W & (cts/ks)& \\
\noalign{\smallskip}\hline \noalign{\smallskip}
\endfirsthead
\caption{continued.}\\
\hline\hline
\noalign{\smallskip}
Name& RA& DEC& I\ \ \ \ & R$-$I& I$-$J& SpT$^{\,a}$& C/& Count rate$^{\,b}$&
$\log L_\mathrm{X}^{\,c}$\ \ \\
&\multicolumn{2}{c}{(J2000)}& & & & & W & (cts/ks)& \\
\noalign{\smallskip}\hline \noalign{\smallskip}
\endhead
\noalign{\smallskip}\hline 
\endfoot
\noalign{\smallskip}\hline \noalign{\smallskip}
\multicolumn{10}{l}{$a$: see note to Table~\ref{detmem}}\\
\multicolumn{10}{l}{$b$: upper limits higher than 10 cts/ks are due to the presence of a very close
bright X-ray source and are therefore }\\
\multicolumn{10}{l}{\phantom{$b$:} overestimated}\\
\multicolumn{10}{l}{$c$: unabsorbed X-ray luminosity in the $0.3-8$ keV band}\\
\endlastfoot
4771-0961       & 5:38:02.21& $-$2:29:55.6& 12.99& 0.69& 1.16& K5      &   & $<$  0.44& $<$ 28.64\\
4771-0706       & 5:38:10.65& $-$2:32:57.4& 12.55& 0.65& 1.12& K4      &   & $<$  0.32& $<$ 28.50\\
r053812-0232    & 5:38:12.60& $-$2:33:01.5& 12.92& 0.73& 0.46& K6      &   & $<$  0.30& $<$ 28.47\\
r053838-0226    & 5:38:38.56& $-$2:26:44.8& 13.33& 0.65& 1.03& K4      &   & $<$  0.37& $<$ 28.56\\
p053834-0239    & 5:38:34.79& $-$2:39:30.0& 11.91& 1.04& 1.47& M1      &   & $<$  0.47& $<$ 28.67\\
p053834-0232    & 5:38:34.85& $-$2:32:52.2& 10.75& 0.54& 0.86& K3      &   & $<$  0.37& $<$ 28.56\\
4771-0740       & 5:38:39.62& $-$2:26:49.7& 12.04& 0.52& 0.82& K2$^*$  &   & $<$  0.37& $<$ 28.56\\
4771-1030       & 5:38:44.80& $-$2:33:57.6& 11.09& 0.67& 1.08& K5      &   & $<$  0.33& $<$ 28.51\\
4771-1090       & 5:38:46.05& $-$2:43:47.8& 13.12& 0.74& 1.15& K6      &   & $<$  0.35& $<$ 28.54\\
4771-1057       & 5:38:48.53& $-$2:44:17.8& 12.25& 0.51& 0.77& K2      &   & $<$  0.38& $<$ 28.57\\
p053854-0240    & 5:38:54.35& $-$2:40:03.0& 16.05& 1.58& 1.74& M3      &   & $<$  0.24& $<$ 28.37\\
4771-0645       & 5:39:09.87& $-$2:25:47.9& 12.11& 0.50& 0.74& K2      &   & $<$  0.49& $<$ 28.68\\
4771-0668       & 5:39:11.21& $-$2:25:49.1& 11.76& 0.57& 0.80& K3      &   & $<$  0.55& $<$ 28.73\\
p053925-0231    & 5:39:25.34& $-$2:31:43.7& 13.75& 0.81& 1.32& K7      &   & $<$  0.41& $<$ 28.61\\
4771-0579       & 5:39:30.43& $-$2:35:07.3& 12.71& 0.74& 1.10& K5      &   & $<$  0.46& $<$ 28.66\\
p053933-0236    & 5:39:33.44& $-$2:36:41.9& 12.73& 0.64& 0.88& K4      &   & $<$  0.48& $<$ 28.67\\
p053938-0233    & 5:39:38.32& $-$2:33:05.4& 12.35& 0.54& 0.79& K3      & C?& $<$  0.77& $<$ 28.88\\
p053939-0231    & 5:39:39.82& $-$2:31:21.8& 12.89& 0.76& 1.05& K8$^*$  & C & $<$  0.74& $<$ 28.86\\
SE 5            & 5:38:14.30& $-$2:25:05.5& 17.52& 1.57& 1.65& M3      &   & $<$  0.43& $<$ 28.63\\
SE 14           & 5:38:23.33& $-$2:25:34.6& 15.69& 1.53& 2.01& M3      &   & $<$  0.42& $<$ 28.62\\
SE 16           & 5:38:23.77& $-$2:22:39.8& 18.76& 2.09& 1.82& M5      &   & $<$  0.75& $<$ 28.87\\
SE 39           & 5:39:02.29& $-$2:23:47.3& 18.01& 1.93& 1.76& M5      &   & $<$  0.46& $<$ 28.66\\
SE 56           & 5:38:49.64& $-$2:45:26.9& 14.85& 1.56& 1.61& M3      &   & $<$  0.37& $<$ 28.56\\
SE 97           & 5:39:27.73& $-$2:28:08.4& 18.48&     &     & M3      &   & $<$  0.55& $<$ 28.73\\
SWW 4           & 5:38:36.88& $-$2:36:43.2& 14.97& 1.74& 1.93& M4      &   & $<$  0.39& $<$ 28.58\\
SWW 7           & 5:39:25.61& $-$2:34:04.2& 14.96& 1.75& 1.76& M4      &   & $<$  0.53& $<$ 28.72\\
SWW 11          & 5:38:37.45& $-$2:50:23.6& 14.66& 1.77& 1.85& M4      &   & $<$  1.48& $<$ 29.16\\
SWW 15          & 5:38:43.87& $-$2:37:06.8& 14.80& 1.79& 1.96& M4      &   & $<$  0.51& $<$ 28.70\\
SWW 25          & 5:38:41.60& $-$2:30:28.9& 14.39& 1.45& 1.55& M3      &   & $<$  0.53& $<$ 28.72\\
SWW 29          & 5:38:47.19& $-$2:34:36.8& 14.64& 1.45&     & M3      &   & $<$  0.35& $<$ 28.54\\
SWW 31          & 5:38:39.03& $-$2:45:32.2& 14.58& 1.34& 1.67& M3      &   & $<$  0.39& $<$ 28.58\\
SWW 40          & 5:38:18.25& $-$2:48:14.3& 14.15& 1.39& 1.39& M3      &   & $<$  2.11& $<$ 29.32\\
SWW 45          & 5:39:26.77& $-$2:42:58.3& 15.52& 1.73& 2.34& M5      &   & $<$  0.68& $<$ 28.83\\
SWW 48          & 5:38:42.28& $-$2:37:14.8& 13.06& 0.98& 1.29& M1      &   & $<$ 63.91& $<$ 30.80\\
SWW 50          & 5:38:31.41& $-$2:36:33.8& 13.80& 1.29& 1.63& M2      & C?& $<$  0.31& $<$ 28.48\\
SWW 53          & 5:37:58.40& $-$2:41:26.2& 15.36& 1.84& 2.07& M4$^*$  &   & $<$  0.67& $<$ 28.82\\
SWW 71          & 5:38:55.44& $-$2:41:29.7& 13.59& 1.39& 1.42& M3      &   & $<$  0.24& $<$ 28.37\\
SWW 77          & 5:39:02.98& $-$2:41:27.2& 14.11& 1.29& 1.67& M2      & W?& $<$  0.47& $<$ 28.67\\
SWW 79          & 5:38:39.65& $-$2:30:21.1& 14.85& 2.42&     & M5      &   & $<$ 16.11& $<$ 30.20\\
SWW 86          & 5:38:40.54& $-$2:33:27.6& 14.54& 1.65& 1.74& M4      &   & $<$  0.28& $<$ 28.44\\
SWW 88          & 5:38:05.67& $-$2:40:19.4& 14.07& 1.27& 1.30& M2      &   & $<$  0.60& $<$ 28.77\\
SWW 98          & 5:38:13.16& $-$2:45:51.0& 13.47& 1.18& 1.40& M2      & C?& $<$  0.47& $<$ 28.67\\
SWW 103         & 5:38:23.08& $-$2:36:49.4& 15.72& 1.42& 1.92& M4      &   & $<$  0.26& $<$ 28.41\\
SWW 122         & 5:39:03.57& $-$2:46:27.0& 14.37& 1.45& 1.53& M3      & W?& $<$  0.41& $<$ 28.61\\
SWW 129         & 5:39:08.78& $-$2:31:11.5& 14.97& 1.63& 1.93& M4      &   & $<$  0.33& $<$ 28.51\\
SWW 174         & 5:37:54.86& $-$2:41:09.2& 15.40& 1.69& 1.90& M4      &   & $<$  0.97& $<$ 28.98\\
SWW 181         & 5:37:52.21& $-$2:33:38.0& 14.41& 1.33& 1.48& M2      & W & $<$148.74& $<$ 31.17\\
SWW 188         & 5:38:28.48& $-$2:46:17.0& 15.06& 1.20& 1.24& M2      &   & $<$  0.63& $<$ 28.79\\
SWW 226         & 5:38:18.17& $-$2:43:34.8& 16.00& 1.52& 1.32& M3      &   & $<$  0.38& $<$ 28.57\\
SWW 230         & 5:38:42.86& $-$2:38:52.5& 14.92& 1.11& 1.24& M2      &   & $<$  0.27& $<$ 28.42\\
SWW 232         & 5:38:19.33& $-$2:32:04.2& 15.99& 1.29&     & M3      &   & $<$  0.27& $<$ 28.42\\
S\,Ori 1        & 5:39:11.83& $-$2:27:41.0& 15.08& 1.70& 1.47& M4      &   & $<$  0.78& $<$ 28.89\\
S\,Ori 2        & 5:39:26.33& $-$2:28:37.7& 15.26& 1.80& 1.76& M4      &   & $<$  0.57& $<$ 28.75\\
S\,Ori 4        & 5:39:39.32& $-$2:32:25.2& 15.23& 2.16& 1.79& M5      &   & $<$  0.70& $<$ 28.84\\
S\,Ori 5        & 5:39:20.24& $-$2:38:25.9& 15.40& 1.86& 1.79& M5      &   & $<$  0.34& $<$ 28.52\\
S\,Ori 7        & 5:39:08.22& $-$2:32:28.4& 15.63& 2.07& 1.83& M4-5    &   & $<$  0.44& $<$ 28.64\\
S\,Ori 8        & 5:39:08.09& $-$2:28:44.8& 15.74& 1.87& 1.60& M4      &   & $<$  0.49& $<$ 28.68\\
S\,Ori 11       & 5:39:44.33& $-$2:33:02.8& 16.39& 2.04& 2.10& M6.0$^*$& W?& $<$  0.98& $<$ 28.98\\
S\,Ori 12       & 5:37:57.46& $-$2:38:44.4& 16.28& 1.87& 2.05& M6.0$^*$& W?& $<$  0.74& $<$ 28.86\\
S\,Ori 14       & 5:39:37.60& $-$2:44:30.5& 16.75& 1.94& 2.37& M6      &   & $<$  0.86& $<$ 28.93\\
S\,Ori 15       & 5:38:48.10& $-$2:28:53.6& 16.39& 1.97& 1.92& M5.5$^*$& W?& $<$  0.38& $<$ 28.57\\
S\,Ori 16       & 5:39:15.10& $-$2:40:47.6& 16.62& 1.91& 1.95& M5      &   & $<$  0.30& $<$ 28.47\\
S\,Ori 17       & 5:39:04.49& $-$2:38:35.4& 16.95& 1.88& 2.18& M6.0$^*$& W?& $<$  0.28& $<$ 28.44\\
S\,Ori 18       & 5:38:25.68& $-$2:31:21.7& 16.61& 1.91& 1.94& M5      &   & $<$  0.27& $<$ 28.42\\
S\,Ori 21       & 5:39:34.33& $-$2:38:46.9& 17.25& 2.02& 2.43& M6      &   & $<$  0.50& $<$ 28.69\\
S\,Ori 22       & 5:38:35.36& $-$2:25:22.2& 16.88& 2.04& 2.24& M6.0$^*$& W?& $<$  0.49& $<$ 28.68\\
S\,Ori 27       & 5:38:17.42& $-$2:40:24.3& 17.03& 2.16& 2.20& M7.0$^*$& W & $<$  0.30& $<$ 28.47\\
S\,Ori 28       & 5:39:23.19& $-$2:46:55.8& 17.14& 2.05& 1.81& M5      &   & $<$  0.71& $<$ 28.84\\
S\,Ori 29       & 5:38:29.61& $-$2:25:14.2& 17.02& 2.00& 2.18& M6.5$^*$& C?& $<$  0.41& $<$ 28.61\\
S\,Ori 30       & 5:39:13.08& $-$2:37:50.9& 17.29& 1.71& 2.05& M6$^*$  & W?& $<$  1.18& $<$ 29.06\\
S\,Ori 31       & 5:38:20.88& $-$2:46:13.3& 17.23& 2.35& 2.04& M7.0$^*$& W?& $<$  0.51& $<$ 28.70\\
S\,Ori 35       & 5:37:55.60& $-$2:33:05.3& 17.61& 2.25& 2.39& M6      &   & $<$  0.99& $<$ 28.99\\
S\,Ori 36       & 5:39:26.85& $-$2:36:56.2& 17.71& 1.88& 2.25& M6      &   & $<$  0.40& $<$ 28.59\\
S\,Ori 39       & 5:38:32.44& $-$2:29:57.3& 17.67& 2.09& 2.23& M6.5$^*$& W?& $<$  0.31& $<$ 28.48\\
S\,Ori 41       & 5:39:38.49& $-$2:31:13.1& 18.44& 2.43& 1.73& M6      &   & $<$  0.65& $<$ 28.81\\
S\,Ori 42       & 5:39:23.41& $-$2:40:57.5& 19.04& 2.47& 2.31& M7.5$^*$& C & $<$  0.37& $<$ 28.56\\
S\,Ori 43       & 5:38:13.95& $-$2:35:01.5& 19.05& 2.38&     & M7      &   & $<$  3.71& $<$ 29.56\\
S\,Ori 44       & 5:38:07.13& $-$2:43:21.0& 19.39& 2.31& 2.16& M7.0$^*$& W?& $<$  0.41& $<$ 28.61\\
S\,Ori 47       & 5:38:14.62& $-$2:40:15.4& 20.49& 2.40& 2.96& L1.5$^*$& C?& $<$  0.28& $<$ 28.44\\
S\,Ori 50       & 5:39:10.80& $-$2:37:15.0& 20.48&     & 2.88& M9.0$^*$& W & $<$  0.40& $<$ 28.60\\
S\,Ori 51       & 5:39:03.20& $-$2:30:20.0& 20.23&     & 2.85& M9.0$^*$& C?& $<$  0.35& $<$ 28.54\\
S\,Ori 53       & 5:38:25.10& $-$2:48:03.0& 21.17&     & 3.28& M9.0$^*$& W & $<$  0.76& $<$ 28.87\\
S\,Ori 58       & 5:39:03.60& $-$2:25:36.0& 21.90&     & 3.30& L0.0$^*$& C?& $<$  0.42& $<$ 28.62\\
S\,Ori 60       & 5:39:37.50& $-$2:30:42.0& 22.76&     & 3.62& L2.0$^*$& C?& $<$  0.77& $<$ 28.88\\
S\,Ori 62       & 5:39:42.10& $-$2:30:31.0& 23.03&     & 3.58& L2.0$^*$& C?& $<$  0.86& $<$ 28.93\\
S\,Ori 65       & 5:38:26.10& $-$2:23:05.0& 23.24&     & 3.34& L3.5$^*$& W & $<$  0.51& $<$ 28.70\\
S\,Ori 69       & 5:39:18.10& $-$2:28:55.0& 23.89&     & 3.63& T0$^*$  &   & $<$  0.56& $<$ 28.74\\
S\,Ori 70       & 5:38:10.10& $-$2:36:26.0& 25.03&     & 4.75& T5.5$^*$&   & $<$  0.34& $<$ 28.52\\
S\,Ori 71       & 5:39:00.20& $-$2:37:06.0& 20.02&     & 2.69& L0$^*$  & C & $<$  0.24& $<$ 28.37\\
J053805.5-023557& 5:38:05.52& $-$2:35:57.1& 17.66& 2.10& 2.38& M6      &   & $<$  0.36& $<$ 28.55\\
J053811.9-024557& 5:38:11.90& $-$2:45:56.8& 15.72& 1.87& 1.80& M5      &   & $<$  0.48& $<$ 28.67\\
J053816.0-023805& 5:38:16.10& $-$2:38:04.9& 15.19& 1.57& 1.61& M4      &   & $<$  0.81& $<$ 28.90\\
J053820.5-023409& 5:38:20.50& $-$2:34:09.0& 14.36& 1.71& 1.71& M4.0$^*$& C?& $<$  0.33& $<$ 28.51\\
J053821.3-023336& 5:38:21.38& $-$2:33:36.3& 17.58& 2.13& 2.22& M5      &   & $<$  0.31& $<$ 28.48\\
J053825.4-024241& 5:38:25.43& $-$2:42:41.3& 16.86& 1.80& 1.98& M4      &   & $<$  0.34& $<$ 28.52\\
J053826.2-024041& 5:38:26.23& $-$2:40:41.3& 16.93& 2.13& 2.02& M8.0$^*$& W?& $<$  0.24& $<$ 28.37\\
J053826.8-022846& 5:38:26.84& $-$2:38:46.0& 16.12& 1.90& 2.01& M5      &   & $<$  0.25& $<$ 28.39\\
J053829.0-024847& 5:38:28.97& $-$2:48:47.3& 16.81&     & 1.99& M6$^*$  & W?& $<$  0.72& $<$ 28.85\\
J053833.9-024508& 5:38:33.88& $-$2:45:07.8& 15.98& 1.89& 1.73& M4      &   & $<$  0.61& $<$ 28.78\\
J053834.5-024109& 5:38:34.60& $-$2:41:08.8& 14.72& 1.46& 1.62& M4      &   & $<$  0.30& $<$ 28.47\\
J053836.7-024414& 5:38:36.69& $-$2:44:13.7& 14.32& 1.76& 1.78& M4      &   & $<$  0.45& $<$ 28.65\\
J053838.6-024157& 5:38:38.59& $-$2:41:55.9& 16.52& 1.82& 1.96& M5      &   & $<$  0.27& $<$ 28.42\\
J053844.4-024037& 5:38:44.48& $-$2:40:37.6& 17.28& 2.32& 2.48& M6      &   & $<$  0.44& $<$ 28.64\\
J053844.4-024030& 5:38:44.49& $-$2:40:30.5& 15.02& 1.49& 1.66& M3      &   & $<$ 48.79& $<$ 30.68\\
J053847.5-022711& 5:38:47.54& $-$2:27:12.0& 14.46& 1.74&     & M5.0$^*$& W & $<$106.85& $<$ 31.02\\
J053848.1-022401& 5:38:48.19& $-$2:44:00.8& 16.13& 1.91& 2.06& M5      &   & $<$  0.38& $<$ 28.57\\
J053849.2-022358& 5:38:49.29& $-$2:23:57.6& 16.27& 1.74& 1.91& M4.0$^*$& C & $<$  0.45& $<$ 28.65\\
J053849.5-024934& 5:38:49.50& $-$2:49:34.0& 20.08&     & 2.80& M7      &   & $<$  0.95& $<$ 28.97\\
J053850.6-024244& 5:38:50.61& $-$2:42:42.9& 15.84& 1.85& 2.00& M5      &   & $<$  0.29& $<$ 28.46\\
J053853.8-024459& 5:38:53.82& $-$2:44:58.9& 17.78& 2.08& 2.33& M6      &   & $<$  0.33& $<$ 28.51\\
J053902.1-023501& 5:39:01.94& $-$2:35:02.9& 16.41&     & 1.96& M4      &   & $<$  0.25& $<$ 28.39\\
J053909.9-022814& 5:39:10.01& $-$2:28:11.6& 16.48& 1.63& 1.88& M5.0$^*$& W?& $<$  0.41& $<$ 28.61\\
J053911.4-023333& 5:39:11.40& $-$2:33:32.8& 16.42& 1.89& 1.97& M5.0$^*$& W?& $<$  0.45& $<$ 28.65\\
J053912.8-022453& 5:39:12.89& $-$2:24:53.8& 19.42&     & 2.61& M6.0$^*$& W & $<$  1.53& $<$ 29.18\\
J053922.2-024552& 5:39:22.25& $-$2:45:52.4& 17.05&     & 1.73& M4      &   & $<$  0.50& $<$ 28.69\\
J053929.4-024636& 5:39:29.40& $-$2:46:36.0& 19.73&     & 2.55& M6      &   & $<$  0.72& $<$ 28.85\\
J053936.4-023626& 5:39:36.40& $-$2:36:26.0& 18.46&     & 2.52& M6      &   & $<$  0.52& $<$ 28.71\\
J053943.2-022343& 5:39:43.19& $-$2:32:43.3& 14.70& 1.64& 1.67& M4      &   & $<$  1.32& $<$ 29.11\\
K 8             & 5:38:50.78& $-$2:36:26.8& 15.06& 2.00& 1.95& M4      &   & $<$  4.98& $<$ 29.69\\
K 50            & 5:38:51.00& $-$2:49:14.0& 16.79& 1.79& 1.75& M4      &   & $<$  0.98& $<$ 28.98\\
K 62            & 5:37:52.07& $-$2:36:04.7& 17.30& 1.91& 2.16& M5      &   & $<$  0.70& $<$ 28.84\\
K 65            & 5:38:39.76& $-$2:32:20.3& 17.58& 1.89& 2.69& M5      &   & $<$  0.30& $<$ 28.47\\
B 3.01-67       & 5:38:46.85& $-$2:36:43.5& 15.28& 1.88& 2.06& M5      &   & $<$  0.53& $<$ 28.72\\
B 2.03-233      & 5:39:40.58& $-$2:39:12.3& 17.29& 1.60& 1.89& M4      &   & $<$  0.74& $<$ 28.86\\
\end{longtable}
\normalsize

\section{Unidentified X-ray sources}
\begin{longtable}{rccrr}
\caption{\label{unid} X-ray sources with no known counterpart within
$5\arcsec$. Sources $N_X =$ 59 and 130 have been detected on one MOS only.
The column labeled ``Sign.'' indicates the significance of detection} \\
\hline\hline
\noalign{\smallskip}
NX& RA$_X$& DEC$_X$& Sign.& Count rate \\
&\multicolumn{2}{c}{(J2000)}& & (cts/ks)\\
\noalign{\smallskip} \hline \noalign{\smallskip}
\endfirsthead
\caption{continued.} \\
\hline\hline
\noalign{\smallskip}
NX& RA$_X$& DEC$_X$& Sign.& Count rate \\
&\multicolumn{2}{c}{(J2000)}& & (cts/ks)\\
\noalign{\smallskip} \hline \noalign{\smallskip}
\endhead
\noalign{\smallskip} \hline 
\endfoot
\noalign{\smallskip} \hline 
\endlastfoot
  5& 5:38:03.76& -2:27:27.8&   8.2&   1.22$\,\pm\,$0.25\\
 10& 5:38:10.83& -2:35:59.6&   7.3&   0.65$\,\pm\,$0.14\\
 12& 5:38:13.86& -2:35:09.1&  23.2&   3.71$\,\pm\,$0.30\\
 13& 5:38:14.36& -2:32:39.2&   7.1&   0.78$\,\pm\,$0.16\\
 15& 5:38:16.99& -2:26:10.1&   6.1&   0.93$\,\pm\,$0.22\\
 16& 5:38:17.69& -2:40:00.4&   5.7&   0.39$\,\pm\,$0.11\\
 18& 5:38:17.84& -2:36:53.0&  10.3&   1.01$\,\pm\,$0.16\\
 21& 5:38:21.90& -2:24:25.6&  15.6&   4.33$\,\pm\,$0.51\\
 22& 5:38:22.22& -2:43:48.6&   6.6&   0.65$\,\pm\,$0.15\\
 23& 5:38:22.34& -2:28:50.1&  16.1&   2.30$\,\pm\,$0.26\\
 24& 5:38:22.72& -2:24:48.7&   6.6&   0.81$\,\pm\,$0.23\\
 27& 5:38:24.19& -2:37:10.6&  12.4&   1.18$\,\pm\,$0.16\\
 28& 5:38:25.02& -2:30:57.1&   7.6&   0.74$\,\pm\,$0.15\\
 29& 5:38:25.27& -2:35:36.2&  14.4&   1.49$\,\pm\,$0.17\\
 30& 5:38:26.15& -2:29:09.8&  25.0&   4.18$\,\pm\,$0.33\\
 35& 5:38:28.54& -2:32:30.6&   6.4&   0.54$\,\pm\,$0.12\\
 36& 5:38:28.59& -2:23:59.5&   7.9&   1.60$\,\pm\,$0.32\\
 37& 5:38:28.78& -2:32:55.7&   9.0&   0.85$\,\pm\,$0.15\\
 38& 5:38:28.91& -2:46:00.6&   9.6&   1.44$\,\pm\,$0.24\\
 40& 5:38:29.74& -2:25:45.9&  11.8&   1.82$\,\pm\,$0.27\\
 43& 5:38:31.20& -2:44:01.9&  17.9&   2.81$\,\pm\,$0.28\\
 45& 5:38:31.89& -2:40:54.1&   7.1&   0.57$\,\pm\,$0.12\\
 48& 5:38:32.82& -2:23:29.5&   8.1&   1.61$\,\pm\,$0.30\\
 51& 5:38:33.25& -2:26:50.0&   6.4&   0.78$\,\pm\,$0.18\\
 57& 5:38:34.69& -2:39:15.2&   5.8&   0.49$\,\pm\,$0.12\\
 58& 5:38:35.27& -2:34:37.7& 107.9&  35.62$\,\pm\,$0.88\\
 59& 5:38:35.29& -2:44:17.6&   5.2&   0.93$\,\pm\,$0.41\\
 63& 5:38:37.54& -2:42:51.4&   8.0&   0.80$\,\pm\,$0.16\\
 66& 5:38:38.65& -2:30:19.4&  14.2&   1.56$\,\pm\,$0.23\\
 72& 5:38:41.57& -2:28:18.5&  20.7&   3.19$\,\pm\,$0.29\\
 73& 5:38:42.16& -2:43:15.4&  30.5&   5.45$\,\pm\,$0.36\\
 74& 5:38:42.29& -2:42:27.6&   6.9&   0.54$\,\pm\,$0.12\\
 77& 5:38:43.86& -2:34:50.2&   5.6&   0.63$\,\pm\,$0.18\\
 83& 5:38:46.64& -2:43:11.3&   6.5&   0.69$\,\pm\,$0.15\\
 85& 5:38:47.26& -2:44:21.5&   9.0&   1.01$\,\pm\,$0.17\\
 86& 5:38:47.37& -2:28:23.8&   9.8&   0.98$\,\pm\,$0.17\\
 95& 5:38:49.66& -2:40:27.3&  12.5&   1.18$\,\pm\,$0.16\\
 96& 5:38:49.71& -2:25:01.3&  17.4&   3.74$\,\pm\,$0.39\\
 99& 5:38:50.14& -2:36:53.4&   7.7&   0.58$\,\pm\,$0.12\\
107& 5:38:53.28& -2:38:31.6&   7.3&   0.46$\,\pm\,$0.10\\
108& 5:38:53.32& -2:45:50.9&   6.9&   0.92$\,\pm\,$0.19\\
112& 5:38:55.11& -2:27:51.6&   7.8&   1.00$\,\pm\,$0.19\\
113& 5:38:55.16& -2:31:01.5&   7.0&   0.51$\,\pm\,$0.11\\
115& 5:38:58.27& -2:38:51.3&  16.6&   1.82$\,\pm\,$0.19\\
121& 5:39:00.52& -2:40:51.5&   7.0&   0.67$\,\pm\,$0.14\\
127& 5:39:03.19& -2:35:00.8&   7.1&   0.57$\,\pm\,$0.12\\
128& 5:39:03.34& -2:23:03.8&   7.6&   1.61$\,\pm\,$0.32\\
129& 5:39:03.35& -2:47:31.4&   5.5&   0.65$\,\pm\,$0.18\\
130& 5:39:04.91& -2:41:47.5&   5.8&   1.03$\,\pm\,$0.39\\
133& 5:39:05.63& -2:27:22.2&   5.0&   0.69$\,\pm\,$0.18\\
134& 5:39:05.72& -2:43:16.6&   6.4&   0.54$\,\pm\,$0.15\\
135& 5:39:06.90& -2:30:49.7&   6.7&   0.72$\,\pm\,$0.18\\
136& 5:39:07.05& -2:33:41.5&  12.5&   1.31$\,\pm\,$0.24\\
139& 5:39:08.52& -2:44:22.3&   5.7&   0.61$\,\pm\,$0.15\\
141& 5:39:09.56& -2:23:40.9&   6.7&   0.97$\,\pm\,$0.29\\
142& 5:39:10.94& -2:26:35.5&   7.1&   1.23$\,\pm\,$0.26\\
143& 5:39:11.13& -2:36:46.6&   6.7&   0.73$\,\pm\,$0.16\\
146& 5:39:12.64& -2:26:09.2&   5.2&   1.91$\,\pm\,$0.45\\
152& 5:39:16.50& -2:32:37.2&  12.2&   1.83$\,\pm\,$0.24\\
155& 5:39:17.54& -2:33:13.8&  14.0&   2.14$\,\pm\,$0.27\\
158& 5:39:18.90& -2:45:23.2&  12.2&   2.29$\,\pm\,$0.33\\
162& 5:39:23.66& -2:34:48.0&   9.2&   1.47$\,\pm\,$0.25\\
163& 5:39:24.13& -2:28:21.9&  16.0&   3.79$\,\pm\,$0.44\\
166& 5:39:26.00& -2:44:43.7&  10.7&   2.27$\,\pm\,$0.36\\
168& 5:39:29.49& -2:41:02.0&  13.4&   2.26$\,\pm\,$0.32\\
171& 5:39:36.21& -2:28:25.2&   6.5&   2.01$\,\pm\,$0.51\\
\end{longtable}

\end{document}